\documentclass[zpreprint,zbstpl]{./LaTeX/zeus/zeus_paper}
\usepackage[english]{babel}

\newcommand{\ZcoosysB}{%
The ZEUS coordinate system is a right-handed Cartesian system, with the $Z$
axis pointing in the proton beam direction, referred to as the ``forward
direction'', and the $X$ axis pointing left towards the centre of HERA.
The polar angle, $\theta$, is measured with respect to the proton beam
direction. The coordinate origin is at the nominal interaction point.\xspace}

\newcommand{\ZcoosysfnB}{\footnote{\ZcoosysB}}

\newcommand{\Zdetdesc}{%
A detailed description of the ZEUS detector can be found 
elsewhere~\cite{zeus:1993:bluebook}. A brief outline of the 
components most relevant for this analysis is given
below.\xspace}
\newcommand{\Zctddesc}[1]{%
Charged particles are tracked in the central tracking detector (CTD)~\citeCTD,
which operates in a magnetic field of $1.43\Tesla$ provided by a thin 
superconducting coil. The CTD consists of 72~cylindrical drift chamber 
layers, organised in nine~superlayers covering the polar-angle\ZcoosysfnB~region 
\mbox{$15^\circ<\theta<164^\circ$}. The relative transverse-momentum 
resolution for
full-length tracks is $\sigma(p_T)/p_T=0.0058p_T\oplus0.0065\oplus0.0014/p_T$,
with $p_T$ in $\Gev$. The position of the interaction vertex along the beam direction can be reconstructed from the CTD tracks with a resolution of about 1 cm in CC events.\xspace\\}
\newcommand{\Zcaldesc}{%
\\
The high-resolution uranium--scintillator calorimeter (CAL)~\citeCAL consists 
of three parts: the forward (FCAL), the barrel (BCAL) and the rear (RCAL)
calorimeters. Each part is subdivided transversely into towers and
longitudinally into one electromagnetic section (EMC) and either one (in RCAL)
or two (in BCAL and FCAL) hadronic sections (HAC). The smallest subdivision of
the calorimeter is called a cell.  The CAL relative energy resolutions, 
as measured under
test-beam conditions, are $\sigma(E)/E=0.18/\sqrt{E}$ for electrons and
$\sigma(E)/E=0.35/\sqrt{E}$ for hadrons ($E$ in $\Gev$). The timing resolution of the CAL is better than 1 ns for energy deposits greater than 4.5 \gev. The position of the interaction vertex along the beam direction can be reconstructed from the arrival time of energy deposits in FCAL. The resolution is about 9 cm for events with FCAL energy above 25 \gev, improving to about 7 cm for FCAL energy above 100 \gev.\xspace\\}
\chardef\usc=95
\chardef\til=126
\catcode`\@=11 % @ signs are now treated as letters
\DeclareRobustCommand\xdotspace{\futurelet\@let@token\@xdotspace}
\def\@xdotspace{%
  \ifx\@let@token.\else
  \ifx\@let@token\bgroup.\else
  \ifx\@let@token\egroup.\else
  \ifx\@let@token\/.\else
  \ifx\@let@token\ .\else
  \ifx\@let@token~.\else
  \ifx\@let@token!.\else
  \ifx\@let@token,.\else
  \ifx\@let@token:.\else
  \ifx\@let@token;.\else
  \ifx\@let@token?.\else
  \ifx\@let@token/.\else
  \ifx\@let@token'.\else
  \ifx\@let@token).\else
  \ifx\@let@token-.\else
  \ifx\@let@token\@xobeysp.\else
  \ifx\@let@token\space.\else
  \ifx\@let@token\@sptoken.\else
   .\space
   \fi\fi\fi\fi\fi\fi\fi\fi\fi\fi\fi\fi\fi\fi\fi\fi\fi\fi}
\catcode`\@=12 % @ signs are no longer letters

\newcommand{\stru}[2]{%
   \relax\ifmmode\hbox{\vrule height#1 depth#2 width0pt}%
   \else\vrule height#1 depth#2 width0pt\fi}
\newcommand{\Ronum}[1]{\uppercase\expandafter{\romannumeral#1}}
\newcommand{\ronum}[1]{\expandafter{\romannumeral#1}}
\DeclareRobustCommand{\LaTeXZ}{%
  \LaTeX\kern-.05em4\kern-.1em
  {\raisebox{-0.2ex}{$\scriptstyle\text{ZEUS}$}}\xspace}
\DeclareMathAlphabet{\mathbf}{OT1}{cmr}{bx}{sl}
\newcommand{\eVdist}{\kern-0.06667em}

\newcommand{\Gev}{{\text{Ge}\eVdist\text{V\/}}}

\newcommand{\gev}{{\,\text{Ge}\eVdist\text{V\/}}}

\newcommand{\Tesla}{\,\text{T}}

\newcommand{\slashfrac}[2]{%
  \raisebox{0.5ex}{\ensuremath #1}\kern-0.12em/\kern-0.08em
  \raisebox{-.8ex}{\ensuremath #2}}
\newcommand{\sqr}[3]{%
    {\vcenter{\hrule height.#3ex\hbox{\vrule width.#2ex height#1ex
     \kern#1ex\vrule width.#3ex}\hrule height.#2ex}}}

\catcode`\@=11 % @ signs are now treated as letters
\newcommand{\parenbar}{\mathpalette\p@renb@r}
\def\p@renb@r#1#2{\vbox{%
  \ifx#1\scriptscriptstyle \dimen@.7em\dimen@ii.2em\else
  \ifx#1\scriptstyle \dimen@.8em\dimen@ii.25em\else
  \dimen@1em\dimen@ii.4em\fi\fi \offinterlineskip
  \ialign{\hfill##\hfill\cr
    \vbox{\hrule width\dimen@ii}\cr
    \noalign{\vskip-.3ex}%
    \hbox to\dimen@{$\mathchar300\hfil\mathchar301$}\cr
    \noalign{\vskip-.3ex}%
    $#1#2$\cr}}}
\catcode`\@=12 % @ signs are no longer letters

\newcommand{\MSbar}{\hbox{$\overline{\rm MS}$}\xspace}

\newcommand{\IP}{{\rm I$\kern-0.01667em$P}\xspace}

\mathchardef\qsm=63
\mathchardef\pls=43
\mathchardef\mns=512
\mathchardef\plm=518
\mathchardef\eql=61
\mathchardef\smallleft=300
\mathchardef\smallright=301
\mathchardef\les=316
\mathchardef\gre=318
\mathchardef\leq=532
\mathchardef\grq=533
\catcode`\@=11 % @ signs are now treated as letters
\newcounter{pict@width}
\newcounter{pict@height}
\newlength{\pict@scale}
\newcommand{\psfigadd}[4]{%
\setcounter{pict@width}{1*\ratio{#2+\pict@scale/2}{\pict@scale}}
\setcounter{pict@height}{1*\ratio{#3+\pict@scale/2}{\pict@scale}}
\setlength{\unitlength}{\pict@scale}
\hbox to #2{\hspace{-\fill}\begin{picture}(\thepict@width,\thepict@height)
\put(0,0){\psfig{figure=#1,width=#2,height=#3,clip=}}
\SetScale{0.283466457}
\SetWidth{1.763889}
{#4}
\end{picture}}
}
\newcounter{pict@widthfst}
\newcounter{pict@widthscd}
\newcounter{pict@widthtot}
\newcommand{\psfigaddtwo}[7]{%
\setcounter{pict@widthfst}{1*\ratio{#2+\pict@scale/2}{\pict@scale}}
\setcounter{pict@widthscd}{1*\ratio{#2+#4+\pict@scale/2}{\pict@scale}}
\setcounter{pict@widthtot}{1*\ratio{#2+#4+#6+\pict@scale/2}{\pict@scale}}
\setcounter{pict@height}{1*\ratio{#3+\pict@scale/2}{\pict@scale}}
\setlength{\unitlength}{\pict@scale}
\hbox{\hspace{-\fill}\begin{picture}(\thepict@widthtot,\thepict@height)
\put(0,0){\psfig{figure=#1,width=#2,height=#3,clip=}}
\put(\thepict@widthscd,0){\psfig{figure=#5,width=#6,height=#3,clip=}}
\SetScale{0.283466457}
\SetWidth{1.763889}
{#7}
\end{picture}}
}
\newcommand{\psfigror}[4]{%
\setcounter{pict@width}{1*\ratio{#2+\pict@scale/2}{\pict@scale}}
\setcounter{pict@height}{1*\ratio{#3+\pict@scale/2}{\pict@scale}}
\setlength{\unitlength}{\pict@scale}
\hbox{\begin{picture}(\thepict@width,\thepict@height)
\put(0,\thepict@height){\psfig{figure=#1,width=#3,height=#2,clip=,angle=270}}
\SetScale{0.283466457}
\SetWidth{1.763889}
{#4}
\end{picture}}
}
\newcommand{\psfigrol}[4]{%
\setcounter{pict@width}{1*\ratio{#2+\pict@scale/2}{\pict@scale}}
\setcounter{pict@height}{1*\ratio{#3+\pict@scale/2}{\pict@scale}}
\setlength{\unitlength}{\pict@scale}
\hbox{\begin{picture}(\thepict@width,\thepict@height)
\put(0,0){\psfig{figure=#1,width=#3,height=#2,clip=,angle=90}}
\SetScale{0.283466457}
\SetWidth{1.763889}
{#4}
\end{picture}}
}
\catcode`\@=12 % @ signs are no longer letters
\newlength\listtextwidth

\catcode`\@=11 % @ signs are now treated as letters
\newlength{\@tabfninsert}
\newlength{\@tabfnwidth}
\newcommand{\tabfootnote}[2]{%
  \setlength{\@tabfninsert}{0.8em}
  \setlength{\@tabfnwidth}{\textwidth}
  \addtolength{\@tabfnwidth}{-\@tabfninsert}
  \addtolength{\@tabfnwidth}{-0.4em}
  \noindent\makebox[\@tabfninsert][r]{\footnotesize$^{#1}$\hfil}\hfill%
  \parbox[t]{\@tabfnwidth}{\footnotesize #2\hfill}}
\catcode`\@=12 % @ signs are no longer letters

\newcommand{\lumiele}      {$16.4~\rm{pb}^{-1}$}

\newcommand{\lumielepercerr}      {$1.8\%$}

\newcommand{\Nfinalel}          {650}
\newcommand{\Nmcel}             {660}

\newcommand{\PTM}       {P_{T,{\rm miss}}}

\def\citeCTD{{\cite{%
nim:a279:290,*npps:b32:181,*nim:a338:254%
}}\xspace}
\def\citeCAL{{\cite{%
nim:a309:77,*nim:a309:101,*nim:a321:356,*nim:a336:23%
}}\xspace}

\begin{document}
%------------------------------------------------------------------------------
%       Title sheet
%------------------------------------------------------------------------------
\prepnum{DESY--02--064}

\title{
Measurement of high-\boldmath${Q^2}$ charged current\\
cross sections in $\mathbf{e^-p}$ deep inelastic  \\
scattering at HERA
}                                                       
                    
\author{ZEUS Collaboration}

\abstract{
Cross sections for $e^-p$ charged current deep inelastic scattering 
 have been measured at a centre-of-mass energy of $318\gev$ with an integrated 
luminosity of \lumiele~using the ZEUS detector at HERA. 
Differential cross-sections $d\sigma/d Q^2$, $d\sigma/d x$ and $d\sigma/d y$ are presented for $Q^{2}>200\gev^2$. In addition, $d^2 \sigma/d x d Q^2$ was measured in the kinematic range \mbox{$280\gev^2 <Q^2<30\,000 \gev^2$} and \mbox{$0.015<x<0.42$}. The predictions of the Standard Model agree well with the measured cross sections. The mass of the $W$ boson, determined from a fit to $d \sigma/ d Q^2$, is \mbox{$M_W=80.3\pm 2.1\,{\rm (stat.)}\,\pm 1.2\,{\rm(syst.)}\,\pm 1.0\,{\rm(PDF)}\,\gev$}.}

\makezeustitle

\newcommand{\address}{ }                                                                           
\newcommand{\author}{ }         
\pagenumbering{Roman}                                                                              
                                    % this "%"s are for cosmetics only                             
%\begin{document}                                                                                   
                                                   %                                               
\begin{center}                                                                                     
{                      \Large  The ZEUS Collaboration              }                               
\end{center}                                                                                       
  S.~Chekanov,                                                                                     
  D.~Krakauer,                                                                                     
  S.~Magill,                                                                                       
  B.~Musgrave,                                                                                     
  J.~Repond,                                                                                       
  R.~Yoshida\\                                                                                     
 {\it Argonne National Laboratory, Argonne, Illinois 60439-4815}~$^{n}$                            
\par \filbreak                                                                                     
  M.C.K.~Mattingly \\                                                                              
 {\it Andrews University, Berrien Springs, Michigan 49104-0380}                                    
\par \filbreak                                                                                     
  P.~Antonioli,                                                                                    
  G.~Bari,                                                                                         
  M.~Basile,                                                                                       
  L.~Bellagamba,                                                                                   
  D.~Boscherini,                                                                                   
  A.~Bruni,                                                                                        
  G.~Bruni,                                                                                        
  G.~Cara~Romeo,                                                                                   
  L.~Cifarelli,                                                                                    
  F.~Cindolo,                                                                                      
  A.~Contin,                                                                                       
  M.~Corradi,                                                                                      
  S.~De~Pasquale,                                                                                  
  P.~Giusti,                                                                                       
  G.~Iacobucci,                                                                                    
  G.~Levi,                                                                                         
  A.~Margotti,                                                                                     
  R.~Nania,                                                                                        
  F.~Palmonari,                                                                                    
  A.~Pesci,                                                                                        
  G.~Sartorelli,                                                                                   
  A.~Zichichi  \\                                                                                  
  {\it University and INFN Bologna, Bologna, Italy}~$^{e}$
% ...
\newpage                                                                                           
% ...
$^{\    2}$ on leave of absence at University of                                                   
Erlangen-N\"urnberg, Germany\\                                                                     
% ...
$^{  30}$ supported by Argonne National Laboratory, USA \\                                         
% ...
                                                           %                                       
                                                           %                                       
% \par         % if index listing & table fit to 1 page, put gap here                              
\newpage   % alternatively: go to newpage, if page is too small                                    
                                                           %                                       
% \institute_references_start    % do not touch or move this line !                                
                                                           %                                       
\begin{tabular}[h]{rp{14cm}}                                                                       
$^{a}$ &  supported by the Natural Sciences and Engineering Research                               
          Council of Canada (NSERC) \\                                                             
$^{b}$ &  supported by the German Federal Ministry for Education and                               
          Research (BMBF), under contract numbers HZ1GUA 2, HZ1GUB 0, HZ1PDA 5, HZ1VFA 5\\         
$^{c}$ &  supported by the MINERVA Gesellschaft f\"ur Forschung GmbH, the                          
          Israel Science Foundation, the U.S.-Israel Binational Science                            
          Foundation, the Israel Ministry of Science and the Benozyio Center                       
          for High Energy Physics\\                                                                
$^{d}$ &  supported by the German-Israeli Foundation, the Israel Science                           
          Foundation, and by the Israel Ministry of Science\\                                      
$^{e}$ &  supported by the Italian National Institute for Nuclear Physics (INFN) \\                
$^{f}$ &  supported by the Japanese Ministry of Education, Science and                             
          Culture (the Monbusho) and its grants for Scientific Research\\                          
$^{g}$ &  supported by the Korean Ministry of Education and Korea Science                          
          and Engineering Foundation\\                                                             
$^{h}$ &  supported by the Netherlands Foundation for Research on Matter (FOM)\\                   
$^{i}$ &  supported by the Polish State Committee for Scientific Research,                         
          grant no. 620/E-77/SPUB-M/DESY/P-03/DZ 247/2000-2002\\                                   
$^{j}$ &  partially supported by the German Federal Ministry for Education                         
          and Research (BMBF)\\                                                                    
$^{k}$ &  supported by the Fund for Fundamental Research of Russian Ministry                       
          for Science and Edu\-cation and by the German Federal Ministry for                       
          Education and Research (BMBF)\\                                                          
$^{l}$ &  supported by the Spanish Ministry of Education and Science                               
          through funds provided by CICYT\\                                                        
$^{m}$ &  supported by the Particle Physics and Astronomy Research Council, UK\\                   
$^{n}$ &  supported by the US Department of Energy\\                                               
$^{o}$ &  supported by the US National Science Foundation\\                                        
$^{p}$ &  supported by the Polish State Committee for Scientific Research,                         
          grant no. 112/E-356/SPUB-M/DESY/P-03/DZ 301/2000-2002, 2 P03B 13922\\                    
$^{q}$ &  supported by the Polish State Committee for Scientific Research,                         
          grant no. 115/E-343/SPUB-M/DESY/P-03/DZ 121/2001-2002, 2 P03B 07022\\                    
\end{tabular}                                                                                      
                                                           %                                       
% \institute_references_end     % do not touch or move this line !                                 
                                                           %                                       
%\end{document}                                                                                     

%------------------------------------------------------------------------------
%       Text
%------------------------------------------------------------------------------

\pagenumbering{arabic} 
\pagestyle{plain}

%%%%%%%%%%%%%%%%%%%%%%%%%%%%%%%%%%%%%%%%%%%%%%%%%%%%%%%%%%%%%%%%%%%%%%%%
%  --------  Introduction  --------  --------  --------  --------
%%%%%%%%%%%%%%%%%%%%%%%%%%%%%%%%%%%%%%%%%%%%%%%%%%%%%%%%%%%%%%%%%%%%%%%%
%
\section{\bf Introduction}
\label{s:introduction}

  Deep inelastic scattering (DIS) of leptons on nucleons
has been vital in the development of our understanding
of the structure of the nucleon.
In the Standard Model (SM),
charged current (CC) DIS is mediated by the exchange of 
the $W$ boson.
In contrast to neutral current (NC) interactions, where all quark and
antiquark flavours contribute, only up-type quarks
and down-type antiquarks 
participate at leading order in $e^-p$ CC DIS reactions.
This makes such interactions a powerful tool for flavour-specific 
investigation of the parton distribution functions (PDFs). Since only left-handed quarks and right-handed antiquarks contribute
to CC DIS at HERA, the distribution of the electron-quark 
centre-of-mass scattering angle, $\theta^*$, is a sensitive probe of the
chiral structure of the weak interaction.

  Measurements of the CC DIS cross sections at HERA have been
reported previously by the H1\cite{pl:b324:241,*zfp:c67:565,pl:b379:319} and ZEUS\cite{prl:75:1006,zfp:c72:47} collaborations. 
These data extended the kinematic region covered by fixed-target 
neutrino-nucleus scattering experiments \cite{zfp:c25:29,*zfp:c49:187,*zfp:c53:51,*zfp:c62:575}
by about two orders of magnitude in the negative square of the four-momentum transfer, $Q^2$. In addition, the double-differential $e^+ p$ CC DIS cross section, 
$d^2\sigma / dxdQ^2$, where $x$ is the Bjorken scaling variable, was measured for the first time at high $Q^2$
by the HERA collider experiments \cite{epj:c12:411,epj:c13:609}.
The mass of the exchanged boson in the space-like domain, extracted from a fit to the differential cross-section $d\sigma/dQ^2$, was consistent
with the mass of the $W$ boson measured in time-like processes at LEP and at the Tevatron\cite{epj:c15:1}.

This paper presents measurements of the
$e^-p$ CC DIS single-differential cross-sections $ d \sigma/ d  Q^2$, 
$ d \sigma/ d  x$ and  $ d \sigma/ d  y$,
as well as $d^2 \sigma / dxdQ^2$.
The results are compared to the expectations of the SM. 
The measurements are based on \lumiele~of data collected
during the running periods in 1998 and 1999 when HERA 
collided electrons of energy $27.5\gev$ with protons of energy $920\gev$, 
yielding a centre-of-mass energy of $318\gev$. The data represent 
an increase of a factor of 20 in integrated luminosity over the previous ZEUS
 $e^- p$ measurement~\cite{zfp:c72:47}. Cross sections for $e^- p$ CC DIS were reported recently by the H1 collaboration~\cite{epj:c19:269}.

%%%%%%%%%%%%%%%%%%%%%%%%%%%%%%%%%%%%%%%%%%%%%%%%%%%%%%%%%%%%%%%%%%%%%%%%
%  --------  Kinematics and Standard Model --------  --------
%%%%%%%%%%%%%%%%%%%%%%%%%%%%%%%%%%%%%%%%%%%%%%%%%%%%%%%%%%%%%%%%%%%%%%%%
%
\section{\bf Standard Model prediction}
\label{s:KinematicsSM}

The electroweak Born-level CC DIS differential cross section, 
$d^2\sigma^{CC}_{\rm Born}/dxdQ^2$, for the reaction 
$e^{-} p \rightarrow \nu_e X$, with longitudinally unpolarised beams, can be expressed as \cite{ijmp:a13:3385} 

\begin{equation}
\frac{d^{2}\sigma^{CC}_{Born}(e^{-}p)}{dxdQ^{2}} = \frac{G^{2}_{F}}{4\pi x}
\frac{M_{W}^{4}}{(Q^{2}+M_{W}^{2})^{2}}[Y_{+}F_{2}^{CC}(x,Q^{2})-y^{2}F_{L}^
{CC}(x,Q^{2}) + Y_{-}xF_{3}^{CC}(x,Q^{2})]
\label{e:Bornele}
\end{equation}
where $G_F$ is the Fermi constant, $M_W$ is the mass of the $W$ boson, $x$ is the Bjorken scaling variable, $y=Q^2 /xs$ and $Y_{\pm}=1\pm (1-y)^2$. The centre-of-mass energy in the electron-proton collision is given by $\sqrt{s}=2\sqrt{E_{e}E_{p}}$, where $E_{e}$ and $E_{p}$ are the electron and proton beam energies, respectively. The inelasticity $y$ is related to $\theta^*$ by $y=1/2(1-\cos\theta^*)$. The structure functions $F_{2}^{CC}$ and $xF_{3}^{CC}$, at leading order in QCD, may be written in terms of sums and differences of quark and antiquark PDFs. For longitudinally unpolarised beams,
\begin{equation}
F_{2}^{CC} = x[u(x,Q^{2})+c(x,Q^{2})+\bar{d}(x,Q^{2})+\bar{s}(x,Q^{2})], \nonumber
\end{equation}
\begin{equation}
xF_{3}^{CC} = x[u(x,Q^{2})+c(x,Q^{2})-\bar{d}(x,Q^{2})-\bar{s}(x,Q^{2})], \nonumber
\end{equation}
where, for example, the PDF $u(x,Q^{2})$ gives the number density of up quarks with momentum fraction $x$ at a given $Q^2$. The longitudinal structure function, $F_{L}^{CC}$, is zero at leading order in QCD. At next-to-leading-order (NLO) in QCD $F_{L}^{CC}$ is non-zero but gives a negligible contribution to the cross section except at values of $y$ close to 1, where it can be as large as 10\%. Since the top-quark mass is large and the off-diagonal elements of the CKM matrix are small~\cite{epj:c15:1}, the contribution from the third-generation quarks may be ignored in CC DIS at HERA~\cite{katz:2000:hera}. Since the $u$-quark density, which is well constrained by NC DIS data, 
dominates in $e^-p$ CC DIS, the uncertainties coming from the PDFs are small. 
The uncertainty in the prediction for 
$d\sigma/dQ^2$, for example, increases from 2\% to 5\% over the $Q^2$
range of the present measurements. 
In the following, uncertainties in the predicted cross sections were obtained using the ZEUS NLO QCD fit~\cite{epj:c21:443}.

%%%%%%%%%%%%%%%%%%%%%%%%%%%%%%%%%%%%%%%%%%%%%%%%%%%%%%%%%%%%%%%%%%%%%%%%
% --------  Detector chapter --------  --------  --------  --------
%%%%%%%%%%%%%%%%%%%%%%%%%%%%%%%%%%%%%%%%%%%%%%%%%%%%%%%%%%%%%%%%%%%%%%%%
%
\section{\bf The ZEUS experiment}
\label{s:detector}

\Zdetdesc

\Zctddesc

\Zcaldesc

An instrumented-iron backing calorimeter (BAC)~\cite{nim:a313:126} surrounds the CAL and can be used to measure energy leakage and identify muons. Muon chambers in the forward, barrel and rear~\cite{nim:a333:342} regions are used in this analysis to veto background events induced by cosmic-ray or beam-halo muons.

The luminosity was measured using the Bethe-Heitler reaction $ep
\rightarrow e \gamma p$. 
The photons were measured by the luminosity monitor~\cite{desy-92-066,*zfp:c63:391,*acpp:b32:2025}, a lead-scintillator calorimeter placed in the HERA tunnel 107 m from the interaction point in the electron beam direction.

%%%%%%%%%%%%%%%%%%%%%%%%%%%%%%%%%%%%%%%%%%%%%%%%%%%%%%%%%%%%%%%%%%%%%%%
% --------  Monte Carlo Simulation --------  --------  --------  --------
%%%%%%%%%%%%%%%%%%%%%%%%%%%%%%%%%%%%%%%%%%%%%%%%%%%%%%%%%%%%%%%%%%%%%%%%
%
\section{Monte Carlo simulation}
\label{s:MCsimulation}

Monte Carlo (MC) simulation was used to determine the efficiency for 
selecting events and the accuracy of kinematic 
reconstruction, to estimate the $ep$ background rates and to extract
cross sections for the full kinematic region.
A sufficient number of
events was generated to ensure that the statistical uncertainties arising from the MC simulation were negligible compared to those of the data.	
The MC samples were normalised to the total
integrated luminosity of the data.

The ZEUS detector response was simulated with a program based on 
{\sc geant}~3.13~\cite{tech:cern-dd-ee-84-1}.  The generated events were passed through 
the simulated detector, subjected to the same trigger requirements as 
the data, and processed by the same reconstruction programs.

Charged current DIS events, including electroweak radiative effects, were 
simulated using the {\sc heracles} 4.6.1~\cite{cpc:69:155,*spi:www:heracles} 
program with the
{\sc djangoh} 1.1~\cite{spi:www:djangoh11} interface to the MC generators that provide the hadronisation. Corrections for initial-state radiation, vertex and 
propagator corrections and two-boson exchange are included in {\sc heracles}.
The mass of the $W$ boson was calculated using the PDG~\cite{epj:c15:1} values for the fine structure constant, the Fermi constant, the mass of the $Z$ boson and the mass of the top quark, and with the Higgs-boson mass set to $100~\gev$.
The colour-dipole model of {\sc ariadne} 4.10~\cite{cpc:71:15}
was used to simulate the $\mathcal{O}(\alpha_{S})$ plus leading logarithmic corrections to the result of the quark-parton model. As a systematic check, the {\sc meps} model of
{\sc lepto} 6.5~\cite{cpc:101:108} was used.
Both programs use the Lund string model of {\sc jetset} 7.4~\cite{cpc:39:347,*cpc:43:367,*cpc:82:74}
for the hadronisation.
A set of NC events generated with {\sc djangoh} was used to estimate
the NC contamination in the CC sample.
Photoproduction background was estimated using events 
simulated with {\sc herwig} 5.9~\cite{cpc:67:465}.
The background from 
$W$ production was estimated
using the {\sc epvec}~\cite{np:b375:3} generator, and the background from
production of charged-lepton pairs was generated with the {\sc lpair}~\cite{proc:hera:1991:1478} program.

%%%%%%%%%%%%%%%%%%%%%%%%%%%%%%%%%%%%%%%%%%%%%%%%%%%%%%%%%%%%%%%%%%%%%%%%
% --------  Kinematic Reconstruction --------  --------  --------  --------
%%%%%%%%%%%%%%%%%%%%%%%%%%%%%%%%%%%%%%%%%%%%%%%%%%%%%%%%%%%%%%%%%%%%%%%%
%
\section{Reconstruction of kinematic variables}
\label{s:reconstruction}

%\subsection{Event Characteristics}
The principal signature of CC DIS at HERA
is the presence of a large
missing transverse momentum, $\PTM$,
arising from the energetic final-state neutrino that escapes detection. 
The quantity $\PTM$ was calculated from
\begin{equation}
\PTM^2  =  P_X^2 + P_Y^2 = 
  \left( \sum\limits_{i} E_i \sin \theta_i \cos \phi_i \right)^2
+ \left( \sum\limits_{i} E_i \sin \theta_i \sin \phi_i \right)^2, \nonumber
  \label{eq:pt}
\end{equation}
where the sums run over all calorimeter energy deposits, $E_i$ 
(uncorrected in the trigger, but corrected in the
offline analysis for energy loss in inactive material etc.~\cite{epj:c11:427}) and
%with polar angle $\theta_i$ and azimuthal angle $\phi_i$.
$\theta_i$ and $\phi_i$ are the polar and azimuthal angles
of the calorimeter deposits as viewed from the interaction vertex.
The polar angle of the hadronic system, $\gamma_h$, is defined by
$\cos\gamma_h = (\PTM^2 - \delta^2)/(\PTM^2 + \delta^2)$,
%\label{eq:gammah}
%\end{equation}
where
%\begin{equation}
$\delta = \sum ( E_i - E_i \cos \theta_{i} ) 
= \sum (E-P_z)_{i}$.
%\label{eq:delta}
%\end{equation}
In the naive quark-parton model,
$\gamma_h$ is the angle through which the struck quark is scattered.
Finally, the total transverse energy,
$E_T$, is given by
%\begin{equation}
$E_T    = \sum E_i \sin \theta_i$. 
%\label{eq:et}
%\end{equation}

%\subsection{Kinematic Reconstruction}

 The kinematic variables were reconstructed
using the Jacquet-Blondel method \cite{proc:epfacility:1979:391}.
The estimators of $y$, $Q^2$ and $x$ are:
%\begin{equation}
$y_{JB} = \delta/(2E_e)$,
% {\hskip 1cm}
$Q^2_{JB} = \PTM^2/(1-y_{JB})$, and
% {\hskip 1cm}
$x_{JB} = Q^2_{JB}/(sy_{JB})$.
%\end{equation}
%where $E_e$ is the positron beam energy.

%To calculate quantities in Eqs. \ref{eq:pt}, \ref{eq:delta} and \ref{eq:et} offline,
%For the offline determination of $\PTM, \delta$ and $E_T$,
%methods developed and tested for the NC cross section
%determination~\cite{ZNCpaper} are used.
%The calorimeter cells with energy deposits
%are grouped into units called clusters.
%For each cluster, corrections are made for hadronic energy
%loss in inactive material in front of the calorimeter depending on
%the cluster energy and angle.
%The correction algorithm, which is based on MC, has been
%verified using the highly constrained NC events measured in the
%ZEUS detector.
%%The Monte Carlo also reveals that 
%Energetic hadron jets
%in the FCAL direction may produce particles backscattered
%into the BCAL or RCAL (albedo).
%Also, particles may be redirected by the material between
%the interaction point and the calorimeter.
%Such effects, which create biases in the measurement of $\gamma_h$, are
%suppressed by removing low energy clusters
%at polar angles much larger than the calculated value of $\gamma_h$.
%The details of the clustering and corrections
%are described in \cite{ZNCpaper}.

%%%%%%%%%%%%%%%%%%%%%%%%%%%%%%%%%%%%%%%%%%%%%%%%%%%%%%%%%%%%%%%%%%%%%%%%
% --------  Event Selection --------  --------  --------  --------
%%%%%%%%%%%%%%%%%%%%%%%%%%%%%%%%%%%%%%%%%%%%%%%%%%%%%%%%%%%%%%%%%%%%%%%%
%
\section{Event selection}
\label{s:EvSel}

Charged current DIS candidates were selected by requiring a large $\PTM$ and a 
reconstructed event vertex consistent with an $ep$ interaction.
The main sources of background come from NC scattering and high-$E_T$
photoproduction. The energy resolution of the CAL
or the energy that escapes detection
can lead to significant missing transverse momentum.  
Events not from $ep$ collisions, such as beam-gas interactions,
beam-halo muons or cosmic rays can also cause substantial apparent
imbalance in the transverse
momentum and so constitute other sources of background.
The selection criteria described below were imposed
to separate CC events from all backgrounds.

When the current jet lies in the central region of the detector, i.e. $\gamma_h$ is large, tracks in the CTD can be used to reconstruct 
an event vertex, strongly suppressing non-$ep$ backgrounds.
The procedure designed to select these events is described
in Section~\ref{ss:StanEvSel}.
For CC events with small $\gamma_h$, the charged particles from 
the hadronic final state 
are often outside the acceptance of the CTD.
Such events populate the high-$x$ region of the kinematic plane.
The algorithm designed specifically to select such events
is described in Section~\ref{ss:LowGEvSel}.
The events were classified first according to $\gamma_0$, the value of
$\gamma_h$ measured with respect to the nominal interaction point.
Subsequently, the kinematic quantities were recalculated using the
$Z$-coordinate of the event vertex ($Z_{\rm VTX}$) determined
from either CTD tracks or the calorimeter-timing information
discussed in Section~\ref{s:detector}.

In the data-taking period, the HERA proton beam contained a significant number of off-axis protons. Such particles passed through the ZEUS interaction region displaced horizontally by several millimeters from the nominal trajectory and gave rise to an especially pernicious class of off-axis beam-gas interactions. In order to remove this background, several selection requirements had to be made more restrictive than was the case in the previous ZEUS study~\cite{epj:c12:411}.

\subsection{Trigger selection}
\label{ss:Trigger}
ZEUS has a three-level trigger system~\cite{zeus:1993:bluebook}. At the first
level, events were selected using criteria based on the energy, transverse 
energy and missing transverse momentum measured in the calorimeter. Generally, events were triggered with a low threshold on these quantities when a
coincidence 
with CTD tracks from the event vertex was required, while a higher threshold was necessary for events with no CTD 
tracks. The latter class of events has a hadronic final state boosted forward outside the CTD 
acceptance. Typical threshold values were $5~\gev$~($8~\gev$)~in missing 
transverse momentum, or $11.5~\gev$~($21~\gev$)~in transverse energy for events with (without) CTD tracks.

At the second level, timing information from the calorimeter was used to reject events inconsistent with the bunch-crossing time. In addition, the topology of the CAL energy deposits was used to reject background events. 
In particular, since the resolution of the missing transverse momentum is 
better at the second level than at the first level, 
a tighter cut of $6~\gev$~($9~\gev$~for events without CTD tracks) was made.

At the third level, track reconstruction and vertex finding were performed and used to reject candidate events with a vertex inconsistent with the beam interaction envelope. Cuts were applied to calorimeter quantities and reconstructed tracks to reduce beam-gas contamination further.

\subsection{Offline selection based on a CTD vertex}
\label{ss:StanEvSel}

Events with $\gamma_{0}>0.4$ rad
were required to have a vertex reconstructed from CTD tracks
and to satisfy all of the following criteria:
\begin{itemize}
  \item {$| Z_{\rm VTX} | < 50$~cm\\}
     the primary vertex reconstructed from the CTD tracks was required to be within 
     the range consistent with the $ep$ interaction region; 
  \item {$\PTM > 12~\gev$ and $\PTM ' > 10~\gev$\\}
    $\PTM '$ is the missing transverse momentum calculated excluding the 
    FCAL towers closest to the beam hole.  
    The $\PTM '$ cut strongly suppresses 
    beam-gas events while maintaining
    high efficiency for CC events;
  \item {Tracking requirement\\}
    at least one track associated with the event vertex must have
    transverse momentum in excess of $0.2\gev$ and 
    a polar angle in the range $15^\circ$ to $164^\circ$. 
    In order to remove off-axis beam-gas background,
    a cut was also applied in two dimensions on the 
    number of such ``good'' tracks, $N_{\rm good}$, versus the
    total number of tracks, $N_{\rm trks}$~\cite{thesis:tapper:2001};
  \item {Rejection of photoproduction\\}
    $\PTM/E_T > 0.5  $ was required for events with  $\PTM < 30\gev$.
    This cut selected a collimated energy flow, as
    expected from a single scattered quark.
    No $\PTM/E_T$ requirement was imposed on the events with
    $\PTM\ge 30\gev$.
    In addition, the difference between the
    direction of the $(P_X,P_Y)$ vector calculated using CTD tracks and
    that obtained using the CAL was required to be 
    less than 0.5~radians for $\PTM <30\gev$ and less than 2.0~radians 
    for $\PTM \geq 30\gev$;
  \item {Rejection of NC DIS\\}
    NC DIS events in which the scattered electron or jet energy was 
    poorly measured could have a large apparent missing transverse momentum.  
    To identify such events,
    a search for candidate electrons was made using an electron-finding 
    algorithm
    that selects isolated electromagnetic clusters in the 
    CAL~\cite{nim:a365:508}.
    Candidate electron clusters within the CTD acceptance were required to 
    have an energy above $4\gev$ and
    a matching track with momentum larger than 25\% of the cluster energy.
    Clusters with $\theta > 164^\circ$ were required to have a transverse
    momentum exceeding $2\gev$.
    Events with a candidate electron satisfying the above criteria
    and $\delta > 20\gev$ were rejected.
    For contained NC events, $\delta$ peaks at $2E_e = 55\gev$.
    This cut was only applied for events with $\PTM < 30\gev$;
  \item {Rejection of non-$ep$ background\\}
    beam-gas events typically gave CAL times that were
    inconsistent with the bunch-crossing time.  Such events were rejected.
    Muon-finding algorithms based on CAL energy deposits or
    muon-chamber signals were used to reject events produced by cosmic rays
    or muons in the beam halo.
\end{itemize}

\subsection{Offline selection based on a CAL vertex}
\label{ss:LowGEvSel}

For events with $\gamma_0<0.4$ rad, the following criteria were imposed:
\begin{itemize}
  \item {$| Z_{\rm VTX} | < 50$~cm\\}
    $Z_{\rm VTX}$ was reconstructed from the
    measured arrival time of energy deposits in FCAL~\cite{pl:b316:412};
  \item {$\PTM > 25\gev$ and $\PTM ' > 25\gev$\\}
    to remove off-axis beam-gas events, the conditions on 
    missing transverse momentum were tightened, compensating for the 
    relaxation of the requirements on tracks;
  \item {Rejection of non-$ep$ background\\}
    the timing and muon-rejection cuts described
    in Section \ref{ss:StanEvSel} were used.
    A class of background events that were especially
    troublesome in this selection
    branch arose from beam-halo muons that produced a shower inside the FCAL.
    To reduce this background,
    topological cuts on the transverse and longitudinal
    shower shape were imposed; these cuts rejected events in which the energy
    deposits were much more strongly collimated than for
    typical hadronic jets.
\end{itemize}

\subsection{Final event sample}
\label{ss:FidBin}

To restrict the sample to regions where
the resolution in the kinematic variables was good and
the backgrounds small, the requirements $Q^2_{JB}>200\gev^2$
and $y_{JB}<0.9$ were imposed.
All events were visually inspected, and nine cosmic-ray and
halo-muon events were removed.
The final $e^- p$ sample consisted of \Nfinalel~events, to be compared with \Nmcel~predicted by the MC simulation.

The resolution in $Q^2$ is $\sim 25\%$ over the entire 
$Q^2$ range. The resolution in $x$ improves from $\sim 25$\%
at $x=0.01$ to $\sim 10\%$ at $x=0.5$.
The resolution in $y$ is $\sim 13\%$ over the 
entire range.

\begin{figure}
  \begin{center}
    \includegraphics*[width=.85\textwidth]{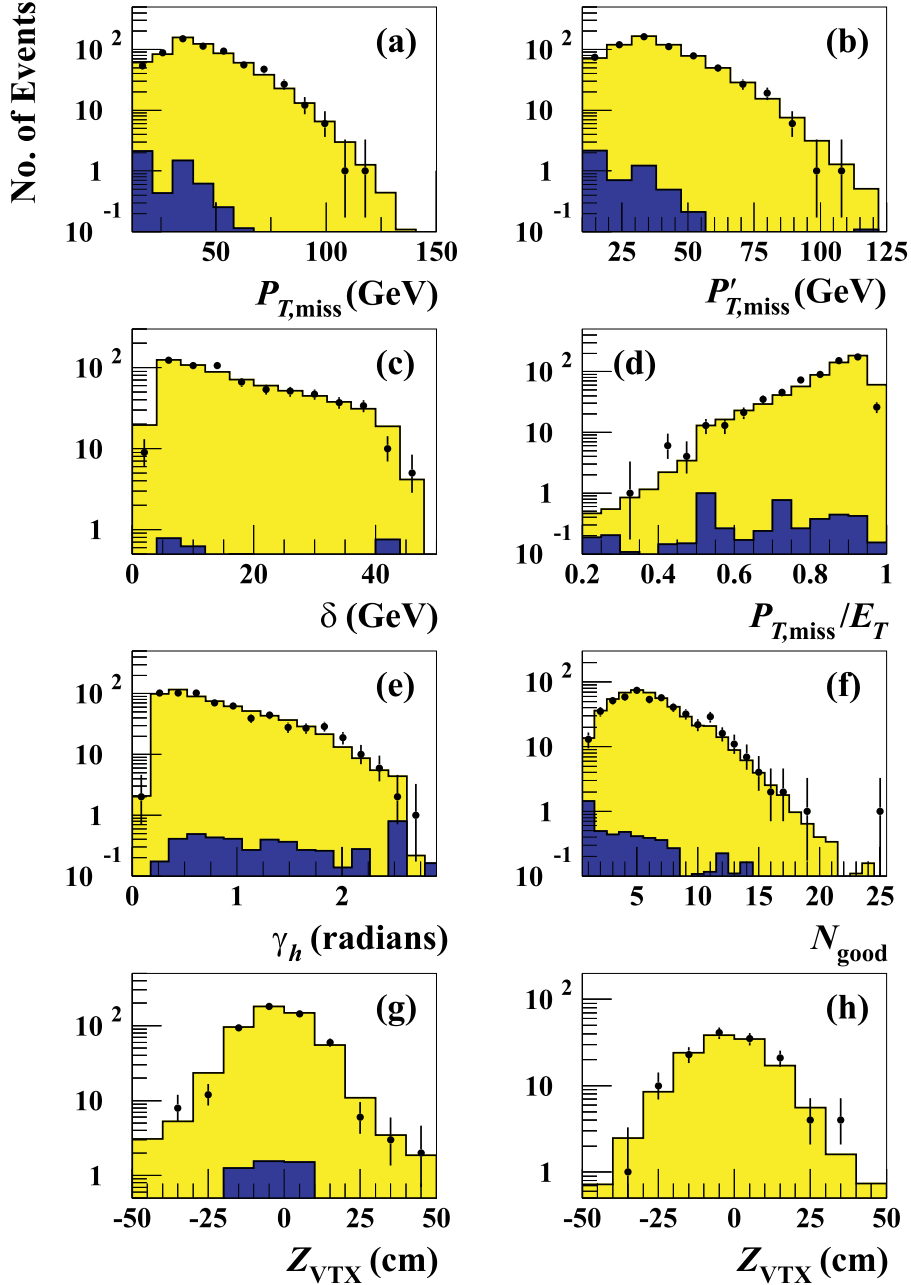}
 \end{center}
  \vskip -10mm
  \caption{
    Comparison of the final $e^- p$ CC data sample (solid points) with the 
    expectations of the sum of the signal and $ep$ background Monte Carlo 
    simulations (light shaded histogram). The $ep$ background Monte Carlo
    is shown as the dark shaded histogram.
    (a) the missing transverse momentum,
    $\PTM$, (b) $\PTM$ excluding the very forward cells, $\PTM'$, 
    (c) the variable 
    $\delta$, defined in the text, (d) 
    the ratio of missing transverse momentum to total transverse energy, 
    $\PTM/E_T$,
    (e) $\gamma_h$, (f) the number of good tracks,
    (g) the $Z$ position of the CTD vertex for the high-$\gamma_0$ sample 
and (h) the $Z$ position of the timing 
    vertex for the low-$\gamma_0$ sample. 
    }
  \label{f:control_el}
\end{figure}

Figure \ref{f:control_el} compares the distributions of data 
events entering the final CC sample with the MC expectation for the sum of the CC signal and $ep$ background events. 
The MC simulation gives a good description of the data.

%%%%%%%%%%%%%%%%%%%%%%%%%%%%%%%%%%%%%%%%%%%%%%%%%%%%%%%%%%%%%%%%%%%%%%
% --------  Method of cross-section determination --------  --------
%%%%%%%%%%%%%%%%%%%%%%%%%%%%%%%%%%%%%%%%%%%%%%%%%%%%%%%%%%%%%%%%%%%%%%%%
%
\section{Cross-section determination and systematic uncertainties}
\subsection{Cross-section determination}
\label{s:xsect}
Monte Carlo events were generated according to Eq.~(\ref{e:Bornele}),
including electroweak radiative effects.
The value of the cross section, at a fixed point within a bin,
was obtained from the ratio of the number of observed events, 
from which the estimated background had been subtracted, to the
number of events predicted by the MC simulation,
multiplied by the cross section obtained using
Eq.~(\ref{e:Bornele}). Consequently, the acceptance, bin-centring and 
radiative corrections were all taken from the MC simulation.

%%%%%%%%%%%%%%%%%%%%%%%%%%%%%%%%%%%%%%%%%%%%%%%%%%%%%%%%%%%%%%%%%%%%%%%%
% --------  Systematic uncertainties --------  --------  --------
%%%%%%%%%%%%%%%%%%%%%%%%%%%%%%%%%%%%%%%%%%%%%%%%%%%%%%%%%%%%%%%%%%%%%%%%
%
\subsection{Systematic uncertainties}
\label{s:systerr}

The major sources of systematic uncertainty
in the cross sections are discussed below:
\begin{itemize}

\item {Uncertainty of the calorimeter energy scale\\}  
  the hadronic energy scale and the associated uncertainty were
  determined, using NC DIS events, from the ratios  of the total hadronic transverse momentum, 
  $P_{T,{\rm had}}$,  to $P_{T,{\rm DA}}$ and $P_{T,{\rm  e }}$, where 
  $P_{T,{\rm DA}} = \sqrt{Q^2_{\rm DA} \left( 1 - y_{\rm DA} \right)}$ 
  is the transverse momentum obtained from the double-angle
  method~\cite{proc:hera:1991:23,*proc:hera:1991:43} and $P_{T,{\rm  e }}$  
  is the measured transverse momentum of the  scattered electron.    
  In order to restrict the hadronic activity to particular polar-angle
  regions, a sample of NC DIS events with a single hadronic jet was selected.
  By applying suitable cuts on the location of the current jet and
  evaluating $P_{T,{\rm had}} / P_{T,{\rm DA}}$ and $P_{T,{\rm had}} / P_{T,{\rm e}}$ 
  event by event, the hadronic energy scales of the FCAL and BCAL were determined.
  The responses of the HAC and EMC sections of the individual
  calorimeters were determined by plotting 
  $P_{T,{\rm had}} / P_{T,{\rm DA}}$ and $P_{T,{\rm had}} / P_{T,{\rm e}}$ 
  as a function of the fraction of the
  hadronic energy measured in the EMC section of the calorimeter. In each case, the 
  uncertainty was found
  by comparing the determinations from data and MC. 
  In order to study the hadronic energy scale in the RCAL, a sample of diffractive DIS 
  events was selected.
  Such events are characterised by a large gap in the hadronic energy
  flow between the proton remnant and the current jet. 
  $P_{T,{\rm had}} / P_{T,{\rm DA}}$  was evaluated event-by-event for events with
  hadronic activity exclusively in the RCAL and the energy scale and associated 
  uncertainty determined.  
 
  The relative uncertainty in the hadronic energy scale was determined to be
  2\% for the RCAL  and 1\% for the FCAL and BCAL.  
  Varying the energy scale of the calorimeter sections by these 
  amounts in the detector simulation induces  small  shifts of the Jacquet-Blondel 
  estimators for the kinematic variables.   
  Varying the energy scale of each of the calorimeters simultaneously up or down
  by these amounts in the detector simulation gives one estimate of the systematic uncertainty. 
  A second estimate was obtained by increasing (decreasing) the FCAL and RCAL energy scales
  together while the BCAL energy scale was decreased (increased). A third estimate was made by
  simultaneously increasing the energy measured in the EMC section of the calorimeter by 2\% and
  decreasing the energy measured in the HAC section by 2\% and vice-versa. 
  This was done separately for each of the calorimeters.	 	 
  The final systematic error attributed to the uncertainty in the hadronic energy
  scale was obtained by taking the quadratic sum of these three estimates.
  The resulting systematic 
  shifts in the measured cross sections were typically within $\pm 5\%$, but increase to $+10\%$ 
  in the highest $y$ bin and $\pm(10-20)\%$ in the highest $Q^{2}$ and $x$ bins;
\item{Energy leakage\\}
  four percent of the accepted events have a measurable energy leakage
  from the CAL into the BAC.
  The average transverse-energy leakage for these events is 5\% of the $E_{T}$   
  observed in the CAL. Both the fraction 
  of events with leakage and the average leakage are well modelled by the MC
  simulation and the effect on the cross-section measurement is negligible;
\item {Variation of selection thresholds\\}
  for the majority of the kinematic bins, varying the selection thresholds 
  by $\pm10\%$ in both data and MC
  resulted in small changes in the measured cross sections. However, varying 
  the $\PTM/E_T$ threshold gave a change in the cross section of around 
  $\pm10\%$ at low $Q^{2}$ in $ d \sigma/ d  Q^2$ and up to $\pm(10-20)\%$ 
  in the bins of the 
  double-differential cross section. 
  Varying the $\PTM '$ cut gave a change of approximately $+10\%$ in the 
  single-differential cross-section $ d \sigma/ d  y$ at the lowest $y$. 
  Varying the $\PTM$ cut gave a change of $-12\%$ in the lowest-$Q^{2}$ bin 
  of $ d \sigma/ d  Q^2$.
  The tracking requirement for $\PTM<30~\gev$~was tightened and gave changes 
  in the cross section of $+8\%$ and $-12\%$ in the lowest-$Q^{2}$ bins of 
  $ d \sigma/ d  Q^2$ and $-10\%$ in the highest $y$ bin of $ d \sigma/ d y$.
  Changes of $\pm(10-20)\%$ were observed in the lowest-$Q^{2}$ 
  bins of the double-differential cross section; 
\item {Uncertainty in the parton-shower scheme\\}
  the {\sc meps} model of {\sc lepto} was used to calculate the acceptance 
  instead of the {\sc ariadne} model.
  The largest effects were observed in the bins of
  low $Q^2$ ($\pm8\%$), highest and lowest $y$ ($\pm5\%$) and
  low $x$ ($\pm4\%$). The largest effect in the double-differential cross 
  section was seen in the low-$Q^2$ and low-$x$ bins, where it amounted to 
  $\pm(10-13)\%$;
\item {Background subtraction\\}
  the uncertainty in the photoproduction background was estimated
  by fitting a linear combination of the $\PTM/E_T$ distributions
  of the signal and the background MC samples to the corresponding
  distribution in the data, allowing the normalisation (N$_{PhP}$) of the 
  photoproduction MC events to vary. 
  No cut on $\PTM/E_T$ was applied for this check. 
  N$_{PhP}$ was varied in the range N$_{PhP}$/2 to $2\cdot$N$_{PhP}$. This 
  range corresponds to the uncertainty given by the fit and 
  changes in the cross sections were found to be within $\pm1\%$;
\item {Trigger efficiency\\}
  the simulation of the efficiency of the first-level trigger as a function
  of $\PTM$ was examined by using data events triggered by an independent 
  trigger branch that was highly efficient for CC events, since it 
  was based on CAL energy sums.
  The difference between the efficiencies calculated from the
  data and from MC events had a negligible effect on the measured cross 
  section;
\item {Choice of parton distribution functions\\} 
  the CC MC events were generated using the CTEQ5D PDFs~\cite{epj:c12:375}.
  The ZEUS NLO QCD fit \cite{epj:c21:443}
  was used to examine the influence of variations of the PDFs on the
  cross-section measurement. Monte Carlo events were re-weighted to reflect the uncertainty from the 
  fit and new acceptance-correction factors were computed.
  The change in the measured cross section was typically $< 1\%$, 
  except at low $Q^{2}$ where it was $-2\%$ and at high $x$ where it 
  was $+3\%$;
\item {The effect of $F_L$\\} 
  the {\sc djangoh} program neglects the $F_L$ contribution to 
  $d^2\sigma/dxdQ^2$ when generating CC events.
  The corresponding 
  effect on the acceptance-correction factors was evaluated by
  re-weighting MC events with the ratio of the cross sections
  with and without $F_L$.  The largest effect was observed in the
  highest-$y$ bin where it was ~$-2\%$;
\item {Uncertainty in the radiative correction\\} 
  the magnitude of the $\mathcal{O}(\alpha)$ electroweak corrections to CC DIS 
  have been discussed by several authors~\cite{jp:g25:1387,proc:mc:1998:530}. 
  Various theoretical approximations and computer codes gave differences in 
  the CC cross sections of typically $\leq~\pm(1-2)\%$. 
  The differences can be as large as ~$\pm(3-8)\%$ at high $x$ and at high $y$.  This uncertainty was not included in the total systematic uncertainty.
\end{itemize}
The individual uncertainties were added in quadrature separately
for the positive and negative deviations from the
nominal cross-section values to obtain the total systematic uncertainty 
listed in Tables \ref{t:xsects} and \ref{t:double}.
The uncertainty on the luminosity of \lumielepercerr~was not included.

%%%%%%%%%%%%%%%%%%%%%%%%%%%%%%%%%%%%%%%%%%%%%%%%%%%%%%%%%%%%%%%%%%%%%%%%
% --------  Cross-section results  --------  --------  --------
%%%%%%%%%%%%%%%%%%%%%%%%%%%%%%%%%%%%%%%%%%%%%%%%%%%%%%%%%%%%%%%%%%%%%%%%
%
\section{Results}
\label{ss:results}

\begin{figure}
  \begin{center}
    \includegraphics[width=.8\textwidth]{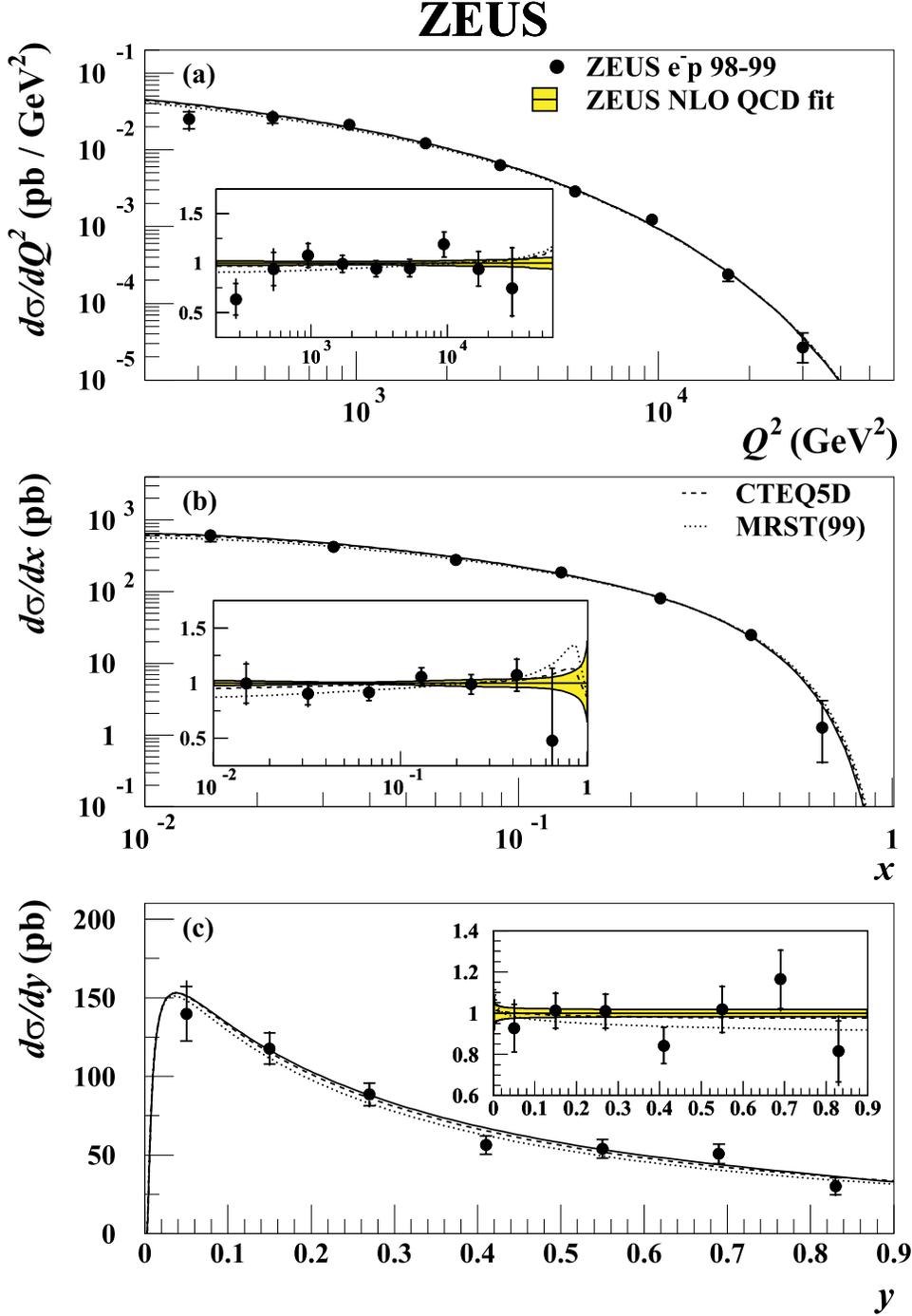}
  \end{center}
  \vskip -2.5mm
  \caption{
    The $e^-p$ CC DIS cross-sections (a) $d\sigma/dQ^2$, (b) $d\sigma/dx$  
    and (c) $d\sigma/dy$ for data (solid points) and the SM expectation 
    evaluated using
    the ZEUS NLO QCD fit, CTEQ5D and MRST(99) PDFs. The insets show
    the ratios of the measured cross sections to the SM expectations 
    evaluated using the ZEUS NLO QCD fit.
    The statistical errors are indicated by the inner error bars (delimited by
    horizontal lines), while the full
    error bars show the total uncertainty obtained by adding the statistical 
    and systematic contributions in quadrature. The shaded band shows the 
    uncertainties associated with the PDFs estimated using the ZEUS NLO QCD fit.
    }
  \label{f:single}
\end{figure}

The single-differential cross-sections $ d \sigma/ d  Q^2$, $ d \sigma/ d  x$ and $ d \sigma/ d  y$ for $Q^{2}>200~\gev^{2}$ are shown in Fig.~\ref{f:single} and compiled\footnote{Tables~\ref{t:uncorr_single} and~\ref{t:uncorr_double} contain details of the systematic uncertainties that are correlated between cross-section bins.} in Table~\ref{t:xsects}. 
The cross sections $ d \sigma/ d  Q^2$
and $ d \sigma/ d  x$ were extrapolated to the full $y$ range using the 
SM predictions with CTEQ5D PDFs.
The SM cross sections derived from Eq.~(\ref{e:Bornele}) using the 
ZEUS NLO QCD fit, CTEQ5D~\cite{epj:c12:375} and 
MRST(99)~\cite{npps:b79:105} 
parameterisations of the PDFs 
are also shown, together with the ratios of the 
measured cross sections to the SM cross section evaluated with the 
ZEUS NLO QCD fit.
The Standard Model gives a good description of the data.

\begin{figure}
  \begin{center}
    \includegraphics[width=0.8\textwidth]{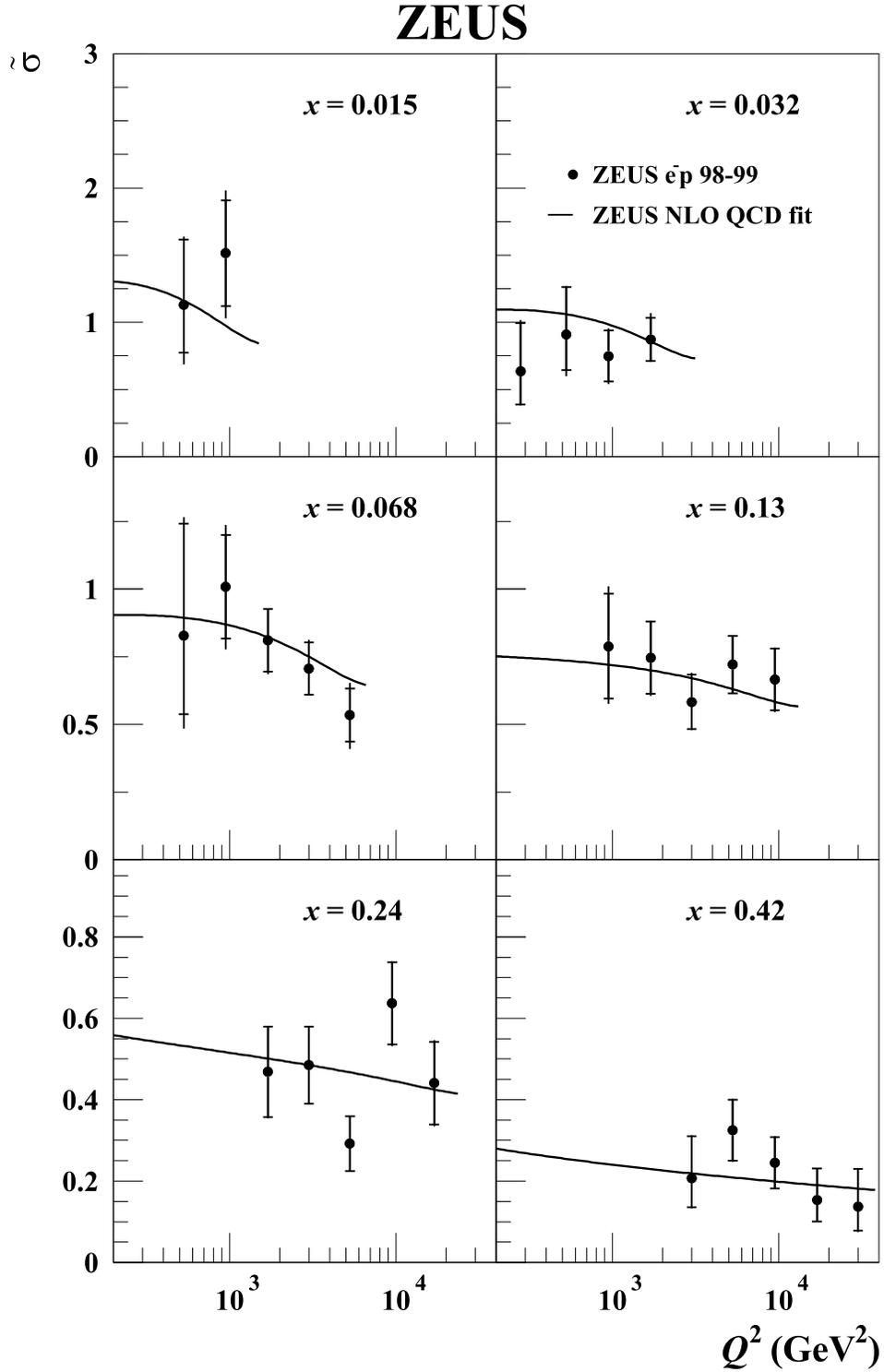}
  \end{center}
  \caption{
    The reduced cross section, $\tilde{\sigma}$, as a function of $Q^2$,
    for different fixed values of $x$.
    The points represent the data, while
    the expectation of the
    Standard Model evaluated using 
    the ZEUS NLO QCD fit is shown as a line.
    }
  \label{f:ddfixx_el}
\end{figure}

\begin{figure}
  \begin{center}
    \includegraphics[width=0.8\textwidth]{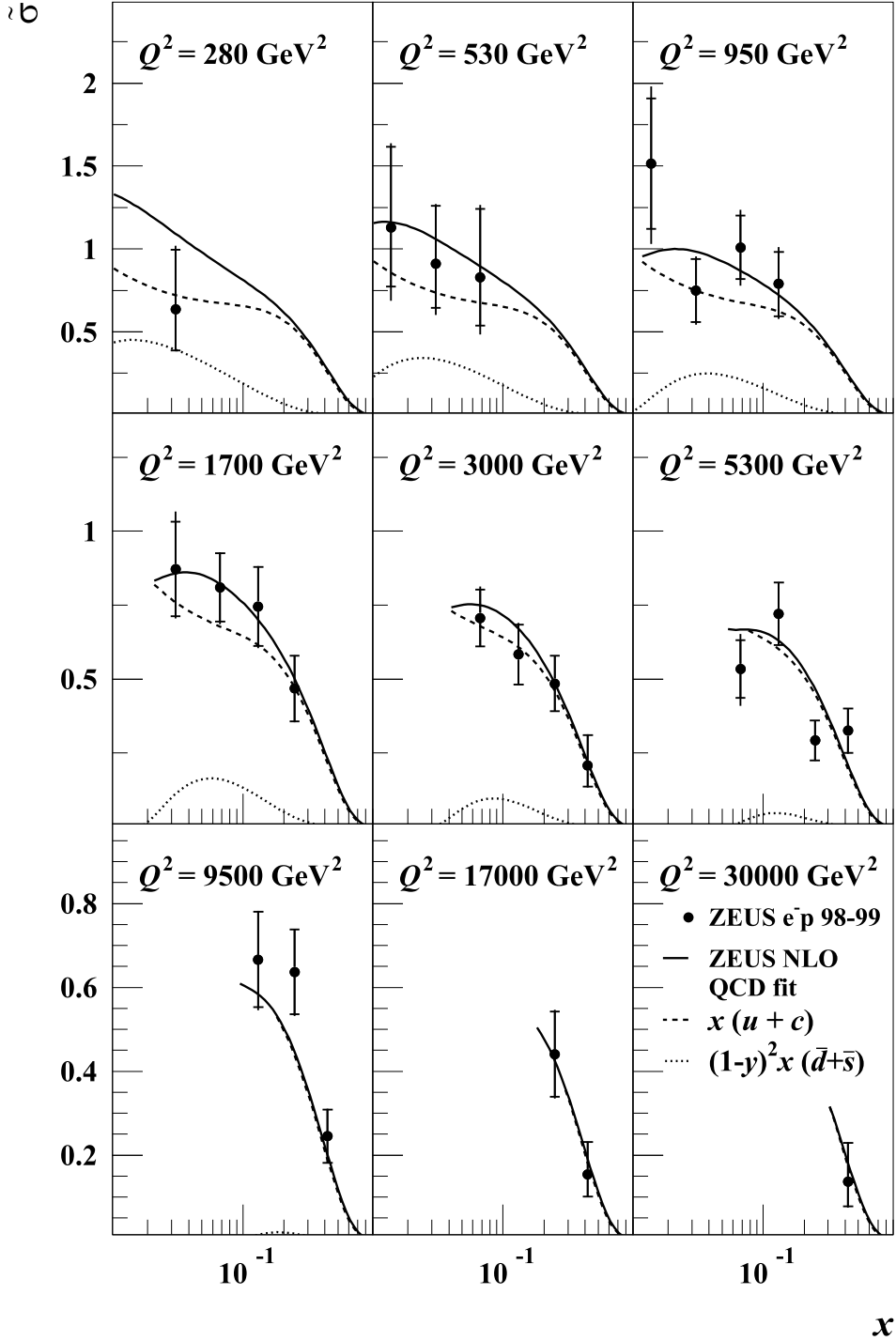}
  \end{center}
  \caption{
    The reduced cross section, $\tilde{\sigma}$, as a function of $x$,
    for different values of $Q^2$.
    The points represent the data, while
    the expectation of the
    Standard Model evaluated using 
    the ZEUS NLO QCD fit
    is shown as a solid line.
    The separate contributions of the PDF combinations $x (u+c)$ and
    $(1-y)^2 x (\bar{d}+\bar{s})$ are shown by the dashed 
    and dotted lines, respectively.
    }
  \label{f:ddfixq2_el}
\end{figure}

The reduced double-differential cross section, $\tilde{\sigma}$, is
defined by
\begin{equation}
   \tilde{\sigma} = \left[{G^2_F \over 2 \pi x } \Biggl( {M^2_W \over M^2_W + Q^2}\Biggr)^2 \right]^{-1}
{{d^2\sigma} \over {dx \, dQ^2}}. \nonumber
  \label{e:Reduced}
\end{equation}
At leading order in QCD, $\tilde{\sigma}({e^- p \rightarrow \nu_e X})$
depends on the quark momentum distributions as follows:
\begin{equation}
  \tilde{\sigma} (e^- p \rightarrow \nu_e X) = x\left[u + c + (1-y)^2 (\bar{d} + \bar{s}) \right].
  \label{e:LO}
\end{equation}
The reduced cross sections are displayed versus $Q^2$ and $x$ 
in Figs. \ref{f:ddfixx_el} and \ref{f:ddfixq2_el}, respectively,  and compiled\footnotemark[2] in Table~\ref{t:double}.
The predictions of Eq.~(\ref{e:Bornele}), evaluated using the
ZEUS NLO QCD fit, give a good description of the data. 
The PDF combinations $(u + c)$ and $(\bar{d} + \bar{s})$, obtained 
in the \MSbar scheme from the ZEUS NLO QCD fit, are shown separately in 
Fig.~\ref{f:ddfixq2_el}. 

The $W$ boson couples only to left-handed fermions and right-handed antifermions. Thus, in the quark-parton-model description of $e^{-}p$ CC DIS, the $Z$ component of the total angular momentum of the initial-state electron-quark system is zero for $e^{-}q$ scattering and one unit of angular momentum for $e^{-}\bar{q}$ scattering. Therefore, the angular distribution of the scattered quark in $e^{-}q$ CC DIS will be flat, while it will exhibit a $(1+\cos\theta^{*})^{2}$ distribution in $e^{-}\bar{q}$ scattering. Since $(1-y)^2 =1/4(1+\cos\theta^{*})^{2}$ the helicity structure of CC interactions 
can be illustrated 
by plotting the reduced double-differential cross section of Eq.~(\ref{e:LO}) 
versus $(1-y)^2$ in bins of fixed $x$.
At leading order in QCD, in the region of approximate scaling, this yields a straight line, the intercept of which gives 
the ($u+c$) contribution, while the slope gives the ($\bar{d}+\bar{s}$) 
contribution.

\begin{figure}
  \begin{center}
    \includegraphics[width=.8\textwidth]{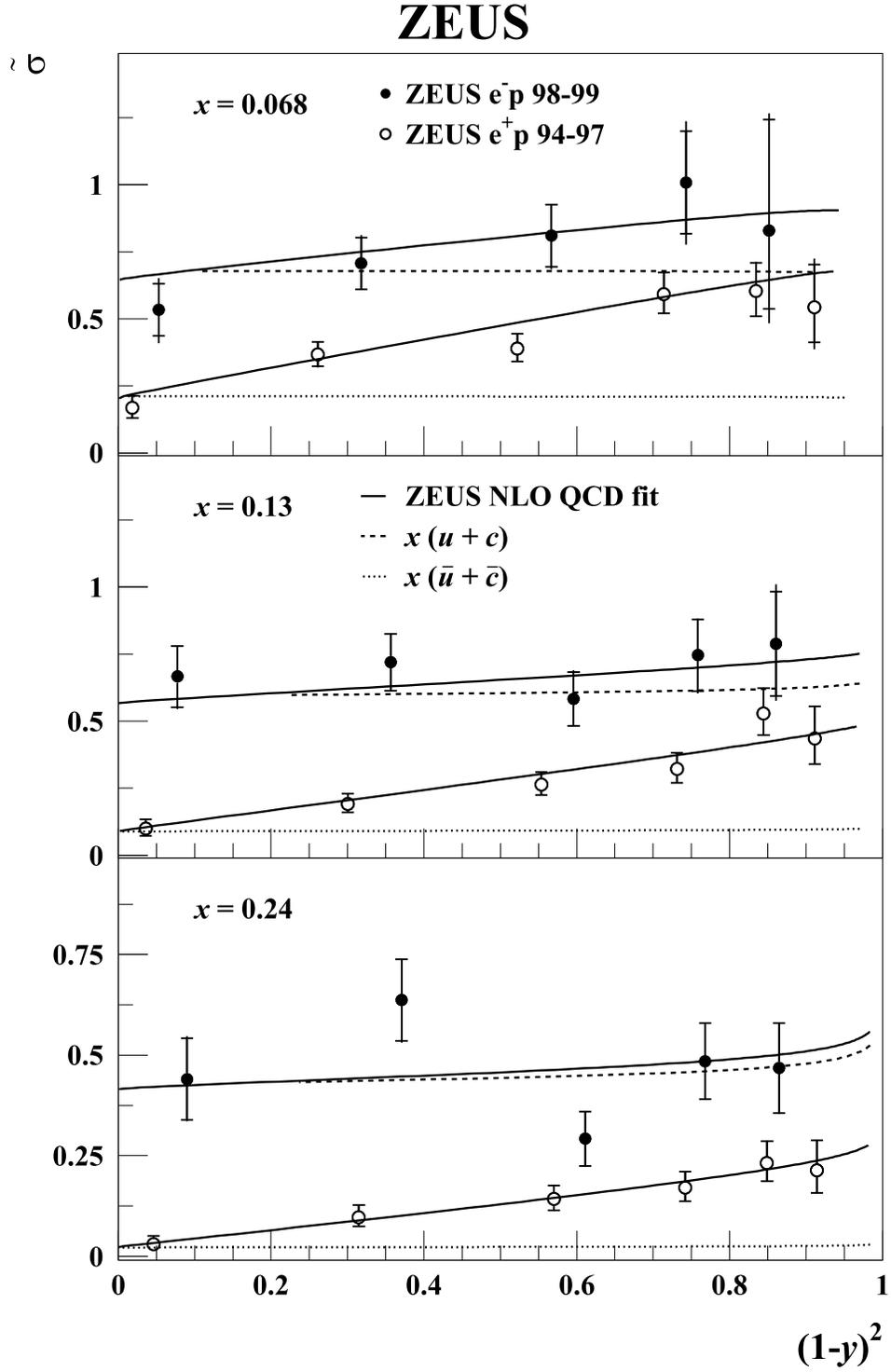}
  \end{center}
  \caption{
    The reduced cross section, $\tilde{\sigma}$, as a function of $(1-y)^2$,
    for different fixed values of $x$, for $e^- p$ (solid points) and $e^+ p$ 
    (open circles) CC DIS.
    The expectation of the
    Standard Model evaluated using 
    the ZEUS NLO QCD fit is shown as a solid line.
    The contributions of the PDF combinations $x (u+c)$ and
    $x (\bar{u}+\bar{c})$ are shown by the dashed 
    and dotted lines, respectively.
    }
  \label{f:helicity_el}
\end{figure}

Figure \ref{f:helicity_el} shows the $e^- p$ CC DIS data, 
compared to the previously published $e^+ p$ data~\cite{epj:c12:411}. 
At large $x$, $e^- p$ CC DIS is sensitive to the $u$-valence-quark PDF while $e^+ p$ CC DIS is sensitive to the $d$-valence-quark PDF. 
At leading order in QCD, the reduced double-differential cross section 
for $e^+ p$ CC DIS can be written as 
\begin{equation}
  \tilde{\sigma}(e^+ p \rightarrow \bar{\nu}_e X) = x\left[\bar{u} + \bar{c} + (1-y)^2 (d + s) \right], \nonumber
\end{equation}
permitting a similar interpretation for the intercept and slope in terms of
the appropriate parton densities. Scaling violations can be observed in the theoretical prediction as $(1-y)^2$ approaches 1.
The data agree with the expectation of the SM from the ZEUS NLO QCD fit, which is also shown. 

The total cross section for $e^- p$ CC DIS in the kinematic region $Q^2 >200\gev^2$ is
\begin{equation}
\sigma^{\rm{CC}}_{\rm{TOT}} (Q^{2} > 200~\rm{GeV}^{2}) = 66.7 \pm 2.7(stat.)^{+1.8}_{-1.1}(syst.)~\rm{pb}. \nonumber
\end{equation}

The result is in good agreement with the SM expectation, evaluated using the ZEUS NLO QCD fit, of $69.0^{+1.6}_{-1.3}$~pb.

%%%%%%%%%%%%%%%%%%%%%%%%%%%%%%%%%%%%%%%%%%%%%%%%%%%%%%%%%%%%%%%%%%%%%%%%
% --------  MW-fit  --------  --------  --------  --------
%%%%%%%%%%%%%%%%%%%%%%%%%%%%%%%%%%%%%%%%%%%%%%%%%%%%%%%%%%%%%%%%%%%%%%%%
%
\section{Electroweak analysis}
\label{ss:MW}

Equation (\ref{e:Bornele}) shows that the magnitude of the 
CC DIS cross section is determined by $G_F$ and the PDFs. The fall 
in the cross section with increasing $Q^2$ is dominated by the propagator 
term, $M_{W}^{4}/(Q^{2}+M_{W}^{2})^{2}$. 
Figure~\ref{f:ddfixx_el} shows that the $Q^2$ dependence of the PDFs is small by comparison. 
An electroweak analysis, performed by fitting 
$ d \sigma/ d  Q^2$ with $G_F$ fixed at the PDG\cite{epj:c15:1} 
value of $1.16639 \cdot 10^{-5}\gev^{-2}$ and $M_W$ treated as a free 
parameter gives
\begin{equation}
M_W=80.3\pm 2.1\,{\rm (stat.)}\,\pm 1.2\,{\rm(syst.)}\,\pm 1.0\,{\rm(PDF)}\,\gev, \nonumber 
\end{equation}

where the third uncertainty was estimated by varying the PDFs within 
the uncertainties given by the ZEUS NLO QCD fit~\cite{epj:c21:443}. 
The systematic uncertainty includes contributions from the sources identified 
in Section \ref{s:systerr} and the uncertainty on the measured 
luminosity.
This measurement, in the space-like region, is in good agreement with 
the more precise measurements of 
$W$-boson production in the time-like region~\cite{epj:c15:1}.
This measurement is also in good agreement with the previous ZEUS measurement of 
\mbox{$M_W=81.4^{+2.7}_{-2.6}{\rm (stat.)}\pm 2.0{\rm(syst.)}^{+3.3}_{-3.0}{\rm(PDF)}\gev$}
obtained by a fit to $ d \sigma/ d  Q^2$ for $e^+ p$ CC DIS at $\sqrt{s}=300$ $\gev$~\cite{epj:c12:411}. The present measurement has a smaller statistical uncertainty than the 
previous result, despite a lower integrated luminosity, due to the larger cross section for CC DIS in $e^- p$ scattering compared to $e^+ p$ scattering. The measurement also benefits from improved understanding of the detector, leading to a smaller systematic uncertainty. Since the $u$-quark density, which is well constrained by NC DIS data, dominates in $e^- p$ CC DIS, the uncertainty coming from the PDFs is smaller than in the case of $e^+ p$ CC DIS, where the $d$-quark density dominates.

%%%%%%%%%%%%%%%%%%%%%%%%%%%%%%%%%%%%%%%%%%%%%%%%%%%%%%%%%%%%%%%%%%%%%%%%
% --------  Summary  --------  --------  --------  --------
%%%%%%%%%%%%%%%%%%%%%%%%%%%%%%%%%%%%%%%%%%%%%%%%%%%%%%%%%%%%%%%%%%%%%%%%
%
\section{Summary}
\label{s:summary}

Differential cross sections for charged current deep inelastic scattering,
\mbox{$e^-p \rightarrow \nu_{e} X$}, have been measured for $Q^{2}>200~\gev^{2}$ 
using \lumiele~of data  collected with the ZEUS detector during the 
period 1998 to 1999. 
The double-differential cross-section $d^2\sigma/dx\,dQ^2$ has been
measured in the kinematic range \mbox{$280\gev^2 <Q^2<30~000 \gev^2$} and \mbox{$0.015<x<0.42$}. The chiral structure of the Standard Model was investigated by plotting the double-differential cross section as a function of $(1-y)^2$. 
The Standard Model gives a good description of the data. The mass of the $W$ boson has been determined, from the $Q^2$ dependence of the measured cross section, to be \mbox{$M_W=80.3\pm 2.1\,{\rm (stat.)}\,\pm 1.2\,{\rm(syst.)}\,\-\pm 1.0\,{\rm(PDF)}\,\gev$}.

%%%%%%%%%%%%%%%%%%%%%%%%%%%%%%%%%%%%%%%%%%%%%%%%%%%%%%%%%%%%%%%%%%%%%%%%
% --------  Acknowledgements  --------  --------  --------  --------
%%%%%%%%%%%%%%%%%%%%%%%%%%%%%%%%%%%%%%%%%%%%%%%%%%%%%%%%%%%%%%%%%%%%%%%%
%
\section*{Acknowledgements}

We appreciate the contributions to the construction and maintenance
of the ZEUS detector of the many people who are not listed as authors.
The HERA machine group and the DESY computing staff are especially
acknowledged for their success in providing excellent operation of the
collider and the data-analysis environment.
We thank the DESY directorate for their strong support and encouragement.

\vfill\eject

%------------------------------------------------------------------------------
%       Bibliography
%------------------------------------------------------------------------------
{
\def\bibname{\Large\bf References}
\def\refname{\Large\bf References}
\pagestyle{plain}
\ifzeusbst
  \bibliographystyle{./BiBTeX/bst/l4z_default}
\fi
\ifzdrftbst
  \bibliographystyle{./BiBTeX/bst/l4z_draft}
\fi
\ifzbstepj
  \bibliographystyle{./BiBTeX/bst/l4z_epj}
\fi
\ifzbstnp
  \bibliographystyle{./BiBTeX/bst/l4z_np}
\fi
\ifzbstpl
  \bibliographystyle{./BiBTeX/bst/l4z_pl}
\fi
{\raggedright
\bibliography{./BiBTeX/user/syn.bib,%
              ./BiBTeX/bib/l4z_articles.bib,%
              ./BiBTeX/bib/l4z_books.bib,%
              ./BiBTeX/bib/l4z_conferences.bib,%
              ./BiBTeX/bib/l4z_h1.bib,%
              ./BiBTeX/bib/l4z_misc.bib,%
              ./BiBTeX/bib/l4z_old.bib,%
              ./BiBTeX/bib/l4z_preprints.bib,%
              ./BiBTeX/bib/l4z_replaced.bib,%
              ./BiBTeX/bib/l4z_temporary.bib,%
              ./BiBTeX/bib/l4z_zeus.bib}}
}
\vfill\eject

%------------------------------------------------------------------------------
%       Tables
%------------------------------------------------------------------------------
%-------------------------------------------------------------------------------

%       An example table

%------------------------------------------------------------------------------

\begin{table}[p]

\begin{center}

\begin{tabular}{|c|c|c|c|c|}

\hline

\multicolumn{5}{|c|}{$d\sigma/dQ^{2}$}\\

\hline

$Q^2$ range ($\gev^2$) & $Q^2$ ($\gev^2$) & N$_{\rm{DATA}}$ & N$_{\rm{BG}}$ & $d\sigma/dQ^{2}$ (pb/$\gev^2$)\\

\hline
 200 - 400      & 280   &       18&       1.0&  $(2.51 \pm 0.63         ^{+0.53}_{-0.46}) \cdot 10^{-2}$\\
 400 - 711      & 530   &       33&       0.3&  $(2.70 \pm 0.48         ^{+0.35}_{-0.44}) \cdot 10^{-2}$\\
 711 - 1265     & 950   &       83&       1.1&  $(2.13 \pm 0.24         ^{+0.13}_{-0.13}) \cdot 10^{-2}$\\
 1265 - 2249    & 1700  &      134&       0.8&  $(1.22 \pm 0.11         ^{+0.04}_{-0.04}) \cdot 10^{-2}$\\
 2249 - 4000    & 3000  &      138&       0.7&  $(6.32 \pm 0.55         ^{+0.13}_{-0.17}) \cdot 10^{-3}$\\
 4000 - 7113    & 5300  &      117&       0.5&  $(2.88 \pm 0.27         ^{+0.10}_{-0.11}) \cdot 10^{-3}$\\
 7113 - 12649   & 9500  &       90&       0.3&  $(1.24 \pm 0.13         ^{+0.04}_{-0.05}) \cdot 10^{-3}$\\
 12649 - 22494  & 17000 &       30&       0.2&  $(2.37 \pm 0.44         ^{+0.16}_{-0.13}) \cdot 10^{-4}$\\
 22494 - 60000  & 30000 &        7&       0.1&  $(2.68 ^{+1.46}_{-1.00}$ $^{+0.30}_{-0.36}) \cdot 10^{-5}$\\
\hline

\multicolumn{5}{|c|}{$d\sigma/dx$}\\

\hline

$x$ range & $x$ & N$_{\rm{DATA}}$ & N$_{\rm{BG}}$ & $d\sigma/dx$ (pb)\\

\hline

 0.010 - 0.021  & 0.015 &       35&       1.0& $(6.08 \pm 1.08          ^{+0.61}_{-0.56}) \cdot 10^{2}$\\

 0.021 - 0.046  & 0.032 &       81&       0.9& $(4.23 \pm 0.48          ^{+0.22}_{-0.33}) \cdot 10^{2}$\\

 0.046 - 0.100  & 0.068 &      172&       1.5& $(2.76 \pm 0.22          ^{+0.05}_{-0.07}) \cdot 10^{2}$\\

 0.100 - 0.178  & 0.130 &      168&       0.7& $(1.87 \pm 0.15          ^{+0.04}_{-0.03}) \cdot 10^{2}$\\

 0.178 - 0.316  & 0.240 &      128&       0.2& $(8.08 \pm 0.72          ^{+0.19}_{-0.17}) \cdot 10^{1}$\\

 0.316 - 0.562  & 0.420 &       55&       0.1& $(2.48 \pm 0.34          ^{+0.15}_{-0.10}) \cdot 10^{1}$\\

 0.562 - 1.000  & 0.650 &        2&       0.1& $(1.28 ^{+1.75}_{-0.86}$ $^{+0.19}_{-0.16}) \cdot 10^{0}$\\

\hline

\multicolumn{5}{|c|}{$d\sigma/dy$}\\

\hline

$y$ range & $y$ & N$_{\rm{DATA}}$ & N$_{\rm{BG}}$ & $d\sigma/dy$ (pb)\\

\hline

 0.00 - 0.10    & 0.05  &       67&       0.6& $(1.40 \pm 0.17          ^{+0.12}_{-0.03}) \cdot 10^{2}$\\

 0.10 - 0.20    & 0.15  &      146&       0.8& $(1.18 \pm 0.10          ^{+0.05}_{-0.05}) \cdot 10^{2}$\\

 0.20 - 0.34    & 0.27  &      153&       0.6& $(8.85 \pm 0.73          ^{+0.26}_{-0.43}) \cdot 10^{1}$\\

 0.34 - 0.48    & 0.41  &       95&       0.7& $(5.63 \pm 0.59          ^{+0.20}_{-0.16}) \cdot 10^{1}$\\

 0.48 - 0.62    & 0.55  &       86&       0.9& $(5.40 \pm 0.59          ^{+0.18}_{-0.19}) \cdot 10^{1}$\\

 0.62 - 0.76    & 0.69  &       70&       0.4& $(5.07 \pm 0.62          ^{+0.19}_{-0.30}) \cdot 10^{1}$\\

 0.76 - 0.90    & 0.83  &       33&       1.0& $(3.01 \pm 0.55          ^{+0.35}_{-0.30}) \cdot 10^{1}$\\

\hline

\end{tabular}

\normalsize

\caption{Values of the differential cross-sections $d \sigma /dQ^{2}$, $d \sigma /dx$ and $d \sigma /dy$. The following quantities are given: the range of the measurement; the value at which the cross section is quoted; the number of data events, $N_{\rm{DATA}}$; the number of expected background events, $N_{\rm{BG}}$ and the measured cross section, with statistical and systematic uncertainties.}

  \label{t:xsects}

\end{center}

\end{table}

\begin{table}[p]

\begin{center}

\begin{tabular}{|c|c|c|c|c|}

\hline

$Q^2$ ($\gev^2$) & $x$ &  N$_{\rm{DATA}}$ & N$_{\rm{BG}}$ & $\tilde{\sigma}$\\

\hline

 280    & 0.032 &        7&       0.3& $0.64  ^{+0.36 }_{-0.25 }$ $     ^{+0.13 }_{-0.06 }$\\

 530    & 0.015 &       10&       0.0& $1.13  ^{+0.49 }_{-0.36 }$ $     ^{+0.14 }_{-0.26 }$\\

 530    & 0.032 &       12&       0.1& $0.91  ^{+0.35 }_{-0.26 }$ $     ^{+0.08 }_{-0.16 }$\\

 530    & 0.068 &        8&       0.0& $0.83  ^{+0.41 }_{-0.29 }$ $     ^{+0.14 }_{-0.19 }$\\

 950    & 0.015 &       17&       0.7& $1.51  \pm 0.39                  ^{+0.25 }_{-0.28 }$\\

 950    & 0.032 &       16&       0.1& $0.75  \pm 0.19                  ^{+0.08 }_{-0.08 }$\\

 950    & 0.068 &       29&       0.2& $1.01  \pm 0.19                  ^{+0.13 }_{-0.13 }$\\

 950    & 0.130 &       17&       0.1& $0.79  \pm 0.19                  ^{+0.11 }_{-0.09 }$\\

 1700   & 0.032 &       31&       0.2& $0.87  \pm 0.16                  ^{+0.11 }_{-0.05 }$\\

 1700   & 0.068 &       51&       0.3& $0.81  \pm 0.12                  ^{+0.02 }_{-0.05 }$\\

 1700   & 0.130 &       32&       0.1& $0.75  \pm 0.13                  ^{+0.03 }_{-0.05 }$\\

 1700   & 0.240 &       18&       0.0& $0.47  \pm 0.11                  ^{+0.01 }_{-0.02 }$\\

 3000   & 0.068 &       56&       0.5& $0.71  \pm 0.10                  ^{+0.04 }_{-0.01 }$\\

 3000   & 0.130 &       34&       0.1& $0.58  \pm 0.10                  ^{+0.03 }_{-0.02 }$\\

 3000   & 0.240 &       27&       0.0& $0.49  \pm 0.09                  ^{+0.01 }_{-0.01 }$\\

 3000   & 0.420 &        8&       0.0& $0.21  ^{+0.10 }_{-0.07 }$ $     ^{+0.02 }_{-0.02 }$\\

 5300   & 0.068 &       31&       0.3& $0.53  \pm 0.10                  ^{+0.07 }_{-0.08 }$\\

 5300   & 0.130 &       47&       0.1& $0.72  \pm 0.11                  ^{+0.02 }_{-0.01 }$\\

 5300   & 0.240 &       19&       0.0& $0.29  \pm 0.07                  ^{+0.01 }_{-0.01 }$\\

 5300   & 0.420 &       19&       0.0& $0.33  \pm 0.07                  ^{+0.02 }_{-0.02 }$\\

 9500   & 0.130 &       35&       0.2& $0.67  \pm 0.11                  ^{+0.03 }_{-0.04 }$\\

 9500   & 0.240 &       40&       0.0& $0.64  \pm 0.10                  ^{+0.02 }_{-0.03 }$\\

 9500   & 0.420 &       15&       0.0& $0.25  \pm 0.06                  ^{+0.02 }_{-0.01 }$\\

 17000  & 0.240 &       19&       0.1& $0.44  \pm 0.10                  ^{+0.03 }_{-0.03 }$\\

 17000  & 0.420 &        8&       0.0& $0.15  ^{+0.08 }_{-0.05 }$ $     ^{+0.01 }_{-0.01 }$\\

 30000  & 0.420 &        5&       0.0& $0.14  ^{+0.09 }_{-0.06 }$ $     ^{+0.02 }_{-0.02 }$\\

\hline

\end{tabular}

\caption{Values of the reduced cross section. The following quantities are given: the values of $Q^2$ and $x$ at which the cross section is quoted; the number of data events, $N_{\rm{DATA}}$; the number of expected background events, $N_{\rm{BG}}$ and the measured cross section, with statistical and systematic uncertainties.}

  \label{t:double}

\end{center}

\end{table}

\begin{table}[p]

\begin{center}

\begin{tabular}{|c|c|c|c||c|c|c|c|}

\hline

\multicolumn{8}{|c|}{$d\sigma/dQ^{2}$}\\

\hline

$Q^2$ ($\gev^2$)     & $d\sigma/dQ^{2}$ (pb/$\gev^2$)& $\delta_{{\rm stat}}$ (\%)&  $\delta_{{\rm sys}}$ (\%)&   $\delta_{{\rm unc}}$ (\%)& $\delta_1$ (\%)& $\delta_2$ (\%)& $\delta_3$ (\%)\\ 

\hline

    280&$2.51 \cdot 10^{-2}$&$ \pm 25$  &$ ^{    +21} _{   -18}$   &$ ^{   +20}_{   -18}$ &$ ^{    +5.1}_{    -0.6}$ &$ ^{    +2.3}_{    +5.5}$ & $ ^{   -3.7}_{   +3.7}$\\
    530&$2.70 \cdot 10^{-2}$&$ \pm 18$  &$ ^{    +13} _{   -16}$   &$ ^{   +10}_{   -13}$ &$ ^{    -1.3}_{    -3.4}$ &$ ^{    -1.5}_{    -4.1}$ & $ ^{   -8.0}_{   +8.0}$\\
    950&$2.13 \cdot 10^{-2}$&$ \pm 11$  &$ ^{    +6.0}_{    -5.9}$ &$ ^{    +4.0}_{    -4.7}$ &$ ^{    +3.4}_{    -1.7}$ &$ ^{    -0.7}_{    +0.4}$ & $ ^{   -3.0}_{   +3.0}$\\
   1700&$1.22 \cdot 10^{-2}$&$ \pm 8.8$ &$ ^{    +3.4}_{    -3.6}$ &$ ^{    +2.5}_{    -2.9}$ &$ ^{    +0.0}_{    -0.3}$ &$ ^{    +0.2}_{    +0.7}$ & $ ^{   -2.1}_{   +2.1}$\\
   3000&$6.32 \cdot 10^{-3}$&$ \pm 8.6$ &$ ^{    +2.0}_{    -2.6}$ &$ ^{    +1.5}_{    -2.5}$ &$ ^{    +0.7}_{    +0.0}$ &$ ^{    +1.0}_{    -0.3}$ & $ ^{   -0.6}_{   +0.6}$\\
   5300&$2.88 \cdot 10^{-3}$&$ \pm 9.3$ &$ ^{    +3.5}_{    -3.7}$ &$ ^{    +1.2}_{    -2.1}$ &$ ^{    -0.7}_{    +1.3}$ &$ ^{    +0.2}_{    +0.2}$ & $ ^{   +3.0}_{   -3.0}$\\
   9500&$1.24 \cdot 10^{-3}$&$ \pm 11$  &$ ^{    +3.3}_{    -4.4}$ &$ ^{    +0.9}_{    -2.8}$ &$ ^{    -3.0}_{    +2.9}$ &$ ^{    +0.3}_{    -1.1}$ & $ ^{   +1.2}_{   -1.2}$\\
  17000&$2.37 \cdot 10^{-4}$&$ \pm 18$  &$ ^{    +6.7}_{    -5.5}$ &$ ^{    +2.6}_{    -1.2}$ &$ ^{    -5.3}_{    +5.9}$ &$ ^{    -0.5}_{    +1.6}$ & $ ^{   +0.4}_{   -0.4}$\\
  30000&$2.68 \cdot 10^{-5}$&$ ^{   +55}_{   -37}$ &$ ^{   +11}_{   -13}$ &$ ^{    +0.8}_{    -5.4}$ &$ ^{   -11}_{   +11}$ &$ ^{    -5.5}_{    +2.6}$ & $ ^{   +1.9}_{   -1.9}$\\

\hline

\multicolumn{8}{|c|}{$d\sigma/dx$}\\

\hline

$x$      &  $d\sigma/dx$ (pb)& $\delta_{{\rm stat}}$ (\%)&  $\delta_{{\rm sys}}$ (\%)&   $\delta_{{\rm unc}}$ (\%)& $\delta_1$ (\%)& $\delta_2$ (\%)& $\delta_3$ (\%)\\ 

\hline

   0.015&$6.08 \cdot 10^2$&$ \pm 18$   &$ ^{    +10} _{    -9.1}$ &$ ^{    +8.5}_{    -9.1}$ &$ ^{    +4.9}_{    -0.9}$ &$ ^{    +2.5}_{    -0.3}$ &$ ^{    -0.1}_{    +0.1}$\\
   0.032&$4.23 \cdot 10^2$&$ \pm 11$   &$ ^{    +5.2}_{    -7.7}$ &$ ^{    +2.6}_{    -6.3}$ &$ ^{    +1.0}_{    -0.7}$ &$ ^{    +0.9}_{    +0.1}$ &$ ^{    -4.3}_{    +4.3}$\\
   0.068&$2.76 \cdot 10^2$&$ \pm 7.8 $ &$ ^{    +1.7}_{    -2.6}$ &$ ^{    +1.6}_{    -2.6}$ &$ ^{    +0.4}_{    -0.4}$ &$ ^{    +0.1}_{    +0.0}$ &$ ^{    -0.3}_{    +0.3}$\\
   0.130&$1.87 \cdot 10^2$&$ \pm 7.8 $ &$ ^{    +2.2}_{    -1.4}$ &$ ^{    +1.2}_{    -0.6}$ &$ ^{    -0.2}_{    +1.2}$ &$ ^{    -0.8}_{    +1.0}$ &$ ^{    +1.0}_{    -1.0}$\\
   0.240&$8.08 \cdot 10^1$&$ \pm 8.9 $ &$ ^{    +2.3}_{    -2.1}$ &$ ^{    +1.6}_{    -1.5}$ &$ ^{    -1.3}_{    +1.3}$ &$ ^{    -0.2}_{    +0.4}$ &$ ^{    -0.8}_{    +0.8}$\\
   0.420&$2.48 \cdot 10^1$&$ \pm 14$   &$ ^{    +5.8}_{    -4.1}$ &$ ^{    +2.1}_{    +0.6}$ &$ ^{    -3.6}_{    +4.9}$ &$ ^{    +2.2}_{    -1.8}$ &$ ^{    -0.2}_{    +0.2}$\\
   0.650&$1.28 \cdot 10^0$&$ ^{  +137}_{   -67}$ &$ ^{   +15}_{   -13}$ &$ ^{    +4.6}_{    -2.6}$ &$ ^{   -12}_{   +13}$ &$ ^{    +5.8}_{    -4.1}$ &$ ^{    +0.8}_{    -0.8}$\\

\hline

\multicolumn{8}{|c|}{$d\sigma/dy$}\\

\hline

$y$      &  $d\sigma/dy$ (pb) & $\delta_{{\rm stat}}$ (\%)&  $\delta_{{\rm sys}}$ (\%)&   $\delta_{{\rm unc}}$ (\%)& $\delta_1$ (\%)& $\delta_2$ (\%)& $\delta_3$ (\%)\\ 

\hline

    0.05&$1.40 \cdot 10^2$&$ \pm 12$  &$ ^{    +8.7}_{    -1.9}$ &$ ^{    +8.6}_{    -1.8}$ &$ ^{    +1.1}_{    -0.4}$ &$ ^{    -0.4}_{    +0.9}$ &$ ^{    +0.1}_{    -0.1}$\\
    0.15&$1.18 \cdot 10^2$&$ \pm 8.4$ &$ ^{    +4.0}_{    -3.9}$ &$ ^{    +1.8}_{    -2.2}$ &$ ^{    +0.6}_{    +0.3}$ &$ ^{    +1.0}_{    +0.5}$ &$ ^{    -3.3}_{    +3.3}$\\
    0.27&$8.85 \cdot 10^1$&$ \pm 8.2$ &$ ^{    +2.9}_{    -4.9}$ &$ ^{    +1.5}_{    -3.3}$ &$ ^{    -0.8}_{    -0.3}$ &$ ^{    -0.2}_{    -2.4}$ &$ ^{    -2.5}_{    +2.5}$\\
    0.41&$5.63 \cdot 10^1$&$ \pm 10$  &$ ^{    +3.6}_{    -2.7}$ &$ ^{    +2.8}_{    -2.6}$ &$ ^{    -0.1}_{    +1.0}$ &$ ^{    +1.9}_{    +0.3}$ &$ ^{    -0.8}_{    +0.8}$\\
    0.55&$5.40 \cdot 10^1$&$ \pm 11$  &$ ^{    +3.3}_{    -3.6}$ &$ ^{    +2.7}_{    -3.0}$ &$ ^{    -0.1}_{    -0.3}$ &$ ^{    +0.3}_{    -0.7}$ &$ ^{    +1.8}_{    -1.8}$\\
    0.69&$5.07 \cdot 10^1$&$ \pm 12$  &$ ^{    +3.7}_{    -5.9}$ &$ ^{    +2.5}_{    -4.7}$ &$ ^{    -1.9}_{    +1.6}$ &$ ^{    -2.3}_{    +1.4}$ &$ ^{    +1.8}_{    -1.8}$\\
    0.83&$3.01 \cdot 10^1$&$ \pm 18$  &$ ^{    +12 }_{    -9.9}$ &$ ^{    +5.8} _{   -8.3}$ &$ ^{    -2.3}_{    +5.7}$ &$ ^{    -1.7}_{    +6.7}$ &$ ^{    +4.6}_{    -4.6}$\\

\hline

\end{tabular}

\normalsize

\caption{Values of the differential cross-sections $d \sigma /dQ^{2}$, $d \sigma /dx$ and $d \sigma /dy$. The following quantities are given: the value at which the cross section is quoted; the measured cross section; the statistical uncertainty; the total systematic uncertainty; the uncorrelated systematic uncertainty and those systematic uncertainties with significant (assumed 100\%) correlations between cross-section bins. The systematic uncertainties considered to be correlated were: the first estimate of the calorimeter energy-scale uncertainty ($\delta_1$); the second such estimate ($\delta_2$) (see text); and the uncertainty in the parton-shower scheme($\delta_3$).}  

\label{t:uncorr_single}

\end{center}

\end{table}

\small
\begin{table}[p]

\begin{center}

\begin{tabular}{|c|c|c|c|c||c|c|c|c|}

\hline

$Q^2$ ($\gev^2$)     & $x$ &$\tilde{\sigma}$ & $\delta_{{\rm stat}}$ (\%)&  $\delta_{{\rm sys}}$ (\%)&   $\delta_{{\rm unc}}$ (\%)& $\delta_1$ (\%)& $\delta_2$ (\%)& $\delta_3$ (\%)\\ 

\hline

    280&   0.032&   0.64&$ ^{   +57}_{   -39}$  &$ ^{    +21}_{    -9.6}$ &$ ^{    +18} _{    -9.3}$&$ ^{    +5.7}_{    +0.1}$&$ ^{    -1.4}_{    +7.1}$&$ ^{    -1.4}_{    +1.4}$\\
    530&   0.015&   1.13&$ ^{   +43}_{   -32}$  &$ ^{    +13}_{   -23}$   &$ ^{    +12} _{    -22}$ &$ ^{    +5.1}_{    -1.2}$&$ ^{    -3.5}_{    -6.8}$&$ ^{    -2.0}_{    +2.0}$\\
    530&   0.032&   0.91&$ ^{   +39}_{   -29}$  &$ ^{    +8.8}_{   -18}$  &$ ^{    +3.1}_{    -15}$ &$ ^{    -1.1}_{    -0.2}$&$ ^{    +2.4}_{    -4.5}$&$ ^{    -7.9}_{    +7.9}$\\
    530&   0.068&   0.83&$ ^{   +50}_{   -35}$  &$ ^{    +17}_{   -22}$   &$ ^{    +11 }_{    -17}$ &$ ^{    -5.7}_{    -4.5}$&$ ^{    -2.5}_{    +1.6}$&$ ^{    -13} _{    +13}$\\
    950&   0.015&   1.51&$ \pm  26             $&$ ^{    +17}_{   -19}$   &$ ^{    +16} _{    -17}$ &$ ^{    +5.9}_{    -5.1}$&$ ^{    -2.8}_{    -4.0}$&$ ^{    -0.8}_{    +0.8}$\\
    950&   0.032&   0.75&$ \pm  25             $&$ ^{    +11}_{   -11}$   &$ ^{    +6.3}_{    -9.7}$&$ ^{    +4.6}_{    -1.1}$&$ ^{    +3.6}_{    +3.8}$&$ ^{    -5.1}_{    +5.1}$\\
    950&   0.068&   1.01&$ \pm  19             $&$ ^{    +12}_{   -13}$   &$ ^{    +9.5}_{    -8.7}$&$ ^{    +0.3}_{    -4.0}$&$ ^{    -2.7}_{    -2.2}$&$ ^{    -8.0}_{    +8.0}$\\
    950&   0.130&   0.79&$ \pm  25             $&$ ^{    +14}_{   -11}$   &$ ^{    +8.0}_{    -4.8}$&$ ^{    +4.2}_{    +3.3}$&$ ^{    -1.9}_{    +2.2}$&$ ^{    +9.6}_{    -9.6}$\\
   1700&   0.032&   0.87&$ \pm  18             $&$ ^{    +13}_{    -5.8}$ &$ ^{    +12} _{    -5.6}$&$ ^{    +0.7}_{    +1.9}$&$ ^{    +2.6}_{    +2.0}$&$ ^{    -1.4}_{    +1.4}$\\
   1700&   0.068&   0.81&$ \pm  14             $&$ ^{    +2.6}_{    -5.9}$&$ ^{    +1.8}_{    -5.4}$&$ ^{    +0.7}_{    -1.0}$&$ ^{    +0.5}_{    -1.4}$&$ ^{    -1.7}_{    +1.7}$\\
   1700&   0.130&   0.75&$ \pm  18             $&$ ^{    +3.5}_{    -6.3}$&$ ^{    +1.3}_{    -4.9}$&$ ^{    -0.3}_{    -1.8}$&$ ^{    -1.3}_{    +0.2}$&$ ^{    -3.3}_{    +3.3}$\\
   1700&   0.240&   0.47&$ \pm  24             $&$ ^{    +2.7}_{    -4.8}$&$ ^{    +1.1}_{    -3.0}$&$ ^{    -1.2}_{    -0.3}$&$ ^{    -2.9}_{    +1.3}$&$ ^{    -2.1}_{    +2.1}$\\
   3000&   0.068&   0.71&$ \pm  14             $&$ ^{    +6.3}_{    -1.9}$&$ ^{    +4.7}_{    -0.3}$&$ ^{    +2.5}_{    +1.6}$&$ ^{    +2.1}_{    +1.1}$&$ ^{    -1.9}_{    +1.9}$\\
   3000&   0.130&   0.58&$ \pm  17             $&$ ^{    +5.9}_{    -3.7}$&$ ^{    +4.9}_{    -2.3}$&$ ^{    +1.0}_{    -0.1}$&$ ^{    +0.9}_{    +0.7}$&$ ^{    +2.9}_{    -2.9}$\\
   3000&   0.240&   0.49&$ \pm  19             $&$ ^{    +1.7}_{    -3.0}$&$ ^{    +0.9}_{    -2.4}$&$ ^{    -0.5}_{    -0.6}$&$ ^{    +0.5}_{    -0.8}$&$ ^{    -1.4}_{    +1.4}$\\
   3000&   0.420&   0.21&$ ^{   +49}_{   -35}$  &$ ^{    +11}_{    -9.8}$ &$ ^{    +8.9}_{    -9.7}$&$ ^{    -0.9}_{    +5.6}$&$ ^{    +2.5}_{    +0.2}$&$ ^{    +1.5}_{    -1.5}$\\
   5300&   0.068&   0.53&$ \pm  18             $&$ ^{    +13}_{   -15}$   &$ ^{    +3.6}_{    -7.5}$&$ ^{    -2.1}_{    +0.1}$&$ ^{    -1.7}_{    +1.1}$&$ ^{    +12}_{     -12}$\\
   5300&   0.130&   0.72&$ \pm  15             $&$ ^{    +2.7}_{    -1.1}$&$ ^{    +1.5}_{    -1.1}$&$ ^{    +0.1}_{    +1.7}$&$ ^{    +0.1}_{    +1.3}$&$ ^{    +0.0}_{    -0.0}$\\
   5300&   0.240&   0.29&$ \pm  23             $&$ ^{    +2.8}_{    -5.1}$&$ ^{    +1.7}_{    -4.9}$&$ ^{    +0.2}_{    +1.6}$&$ ^{    +1.5}_{    -1.2}$&$ ^{    -0.5}_{    +0.5}$\\
   5300&   0.420&   0.33&$ \pm  23             $&$ ^{    +6.0}_{    -6.2}$&$ ^{    +2.3}_{    -2.8}$&$ ^{    -1.0}_{    +2.3}$&$ ^{    +2.5}_{    -3.2}$&$ ^{    -4.4}_{    +4.4}$\\
   9500&   0.130&   0.67&$ \pm  17             $&$ ^{    +3.8}_{    -6.2}$&$ ^{    +1.4}_{    -4.9}$&$ ^{    -3.3}_{    +3.5}$&$ ^{    -2.0}_{    +0.2}$&$ ^{    +0.0}_{    -0.0}$\\
   9500&   0.240&   0.64&$ \pm  16             $&$ ^{    +2.4}_{    -4.7}$&$ ^{    +0.4}_{    -3.6}$&$ ^{    -2.3}_{    +1.1}$&$ ^{    +1.1}_{    -1.1}$&$ ^{    +1.8}_{    -1.8}$\\
   9500&   0.420&   0.25&$ \pm  26             $&$ ^{    +6.3}_{    -5.1}$&$ ^{    +2.5}_{    -0.4}$&$ ^{    -3.5}_{    +3.9}$&$ ^{    +4.1}_{    -3.5}$&$ ^{    -0.9}_{    +0.9}$\\
  17000&   0.240&   0.44&$ \pm  23             $&$ ^{    +7.1}_{    -6.1}$&$ ^{    +2.7}_{    -1.6}$&$ ^{    -5.0}_{    +4.8}$&$ ^{    -3.1}_{    +4.3}$&$ ^{    -0.9}_{    +0.9}$\\
  17000&   0.420&   0.15&$ ^{   +50}_{   -35}$  &$ ^{    +9.6}_{    -7.3}$&$ ^{    +3.1}_{    -1.3}$&$ ^{    -4.9}_{    +6.2}$&$ ^{    +6.2}_{    -4.6}$&$ ^{    +2.3}_{    -2.3}$\\
  30000&   0.420&   0.14&$ ^{   +68}_{   -43}$  &$ ^{   +13}  _{   -13}$  &$ ^{    +1.8}_{    -5.4}$&$ ^{    -9.4}_{    +10}$ &$ ^{    -5.9}_{    +5.1}$&$ ^{    +5.3}_{    -5.3}$\\

\hline

\end{tabular}

\caption{Values of the reduced cross section. The following quantities are given: the values of $Q^2$ and $x$ at which the cross section is quoted; the measured cross section; the statistical uncertainty; the total systematic uncertainty; the uncorrelated systematic uncertainty and those systematic uncertainties with significant (assumed 100\%) correlations between cross-section bins. The systematic uncertainties considered to be correlated were: the first estimate of the calorimeter energy-scale uncertainty ($\delta_1$); the second such estimate ($\delta_2$) (see text); and the uncertainty in the parton-shower scheme($\delta_3$).}

  \label{t:uncorr_double}

\end{center}

\end{table}

%------------------------------------------------------------------------------
%       Figures
%------------------------------------------------------------------------------
%
%--------  Monte Carlo data comparison
%
\newpage
%
%--------  Single Differential Distribution:Electron
%
%
%--------  Double Differential Cross Section:Electron
%

%
%       ... that's it
%
\end{document}